\documentclass[prb,twocolumn,preprintnumbers,amsmath,amssymb,superscriptaddress,longbibliography]{revtex4-1}
\usepackage{graphicx, epsf,bm}% Include figure files
\usepackage{subfigure}
\usepackage{color}
\newcommand{\br}{{\bf r}}
\newcommand{\bR}{{\bf R}}
\newcommand{\bk}{{\bf k}}
\newcommand{\bq}{{\bf q}}

\newcommand{\low}{\mathrm{low}}

\begin{document}
\title{Rare region induced avoided quantum criticality in disordered three-dimensional Dirac and Weyl semimetals}
\author{J. H. Pixley}
\affiliation{Condensed Matter Theory Center and the Joint Quantum Institute, Department of Physics, University of Maryland, College Park, Maryland 20742-4111 USA}
\author{David A. Huse}
\affiliation{Physics Department, Princeton University, Princeton, NJ 08544 USA, and Institute for Advanced Study, Princeton, NJ 08540 USA}
\author{S. Das Sarma}
\affiliation{Condensed Matter Theory Center and the Joint Quantum Institute, Department of Physics, University of Maryland, College Park, Maryland 20742-4111 USA}
\date{\today}

\begin{abstract}
We numerically study the effect of short ranged potential disorder on massless noninteracting three-dimensional Dirac and Weyl fermions,
with a focus on the question of the proposed (and extensively theoretically studied)
quantum critical point separating semimetal and diffusive metal phases.  We determine the properties of the eigenstates of the disordered Dirac
Hamiltonian ($H$) and exactly calculate the density of states (DOS) near zero energy, using a combination of Lanczos on $H^2$ and the
kernel polynomial method on $H$.  We establish the existence of two distinct types of low energy eigenstates contributing to the disordered density of states in the weak disorder semimetal regime.
These are (i) typical eigenstates that are well described by linearly dispersing perturbatively dressed Dirac states, and
(ii) nonperturbative rare eigenstates that are weakly-dispersive and quasi-localized in the real space regions with the largest
(and rarest) local random potential.  Using twisted boundary conditions, we are able to systematically find and study these two (essentially independent) types of eigenstates.
We find that the Dirac states contribute low energy peaks in the finite-size DOS that arise from the clean eigenstates
which shift and broaden
in the presence of disorder. On the other hand, we establish that the rare quasi-localized eigenstates contribute a nonzero
background DOS which is only weakly energy-dependent near zero energy and is exponentially small at weak disorder.
 We also find that the expected semimetal to diffusive metal quantum critical point is converted to an {\it avoided} quantum criticality
that is ``rounded out'' by nonperturbative effects,
with no signs of any singular behavior in the DOS at the energy of the clean Dirac point.
However, the crossover effects of the avoided (or hidden) criticality manifests itself in a so-called quantum critical fan region away from the Dirac energy.
We discuss the implications of our results for disordered Dirac and Weyl semimetals, and reconcile the large body of existing numerical work showing quantum criticality with the existence of these nonperturbative effects.
\end{abstract}

%% Pacs #s
%71.10.Hf	 Non-Fermi-liquid ground states, electron phase diagrams and phase transitions in model systems
%72.80.Ey	III-V and II-VI semiconductors
%73.43.Nq Quantum phase transitions (see also 64.70.Tg Quantum phase transitions in equations of state, phase equilibria and phase transitions)
%72.15.Rn	Localization effects (Anderson or weak localization)

\maketitle

\section{Introduction}
\label{sec:intro}
Recently, there has been an intense experimental effort to find gapless semiconductors that host isolated points in momentum space with linearly touching valence and conduction bands. This thrust has been fueled by the exciting possibility of studying massless three-dimensional Dirac (for Kramers degenerate bands) and Weyl (for non-Kramers degenerate bands) fermions in solid state systems.
(The fact that the two-dimensional version of a Dirac-Weyl system already exists in the form of graphene has obviously been a great impetus in this search for three-dimensional Dirac-Weyl materials.)
This has led to establishing three-dimensional Dirac semimetals in the compounds Cd$_3$As$_2$ (Refs.~\onlinecite{Neupane-2014,Liu-2014,Borisenko-2014}), Na$_3$Bi (Refs.~\onlinecite{Liu2-2014,Xu-2015}), Bi$_{1-x}$Sb$_x$ (Refs.~\onlinecite{Lenoir-1996,Ghosal-2007,Teo-2008}), BiTl(S$_{1-\delta}$Se$_{\delta})_2$ (Refs.~\onlinecite{Xu-2011,Sato-2011}), (Bi$_{1-x}$In$_x$)$_2$Se$_3$ (Refs.~\onlinecite{Brahlek-2012,Wu-2013}), and Pb$_{1-x}$Sn$_x$Te (Refs.~\onlinecite{Dornhaus-1983,Xu-2012,Gibson-2014}). While, even more recently the existence of Weyl semimetals~\cite{Wan-2011,HWeng-2015,Huang-2015} in TaAs (Refs.~\onlinecite{Xu3-2015,Weng-2015}) and NbAs (Ref.~\onlinecite{Xu2-2015})
has been established. This low energy description is also applicable to various other physical systems that host gapless Dirac or Weyl points such as the pyrochlore iridates~\cite{Wan-2011} and the Bugliobov quasiparticle properties of nodal superconductors. With the experimental discovery of such a large number of Dirac-Weyl materials (and the great deal of interest and excitement surrounding them), as established by their electronic band structures through photoemission spectroscopy (i.e. linearly touching conduction and valence bands), one of the immediate important questions is how robust this noninteracting clean system is to the presence of interaction and disorder, physical effects invariably present in real solid state materials.
Here we study the fundamental effects of static potential disorder on noninteracting Dirac-Weyl systems.  (We note that typically these materials are considered to be weakly interacting due to the strong screening provided by the large background lattice dielectric constant in the systems.)

Due to the invariable presence of disorder in all solid state materials there has been a substantial amount of theoretical activity studying the effect of disorder on non-interacting Dirac and Weyl fermions~\cite{Fradkin-1986,Goswami-2011,Kobayashi-2014,Brouwer-2014,Bitan-2014,*Bitan-2016,Nandkishore-2014,Pixley-2015,Altland-2015,Sergey-2015,Leo-2015,Sbierski-2015,Pixley2015disorder,Garttner-2015,Liu-2015,Bera-2015,Shapourian-2015,Altland2-2015,Sergey2-2015}.
Focusing on the undoped (i.e., Fermi energy at $E=0$) Dirac point (i.e. the band touching point), the quadratically vanishing density of states at zero energy ($\rho(E)\sim E^2$) associated with the linear three-dimensional energy band dispersion places these problems in a different class than that of a conventional metal with a parabolic energy dispersion and a nonzero Fermi energy.  In a standard metal, the nonzero density of states at the Fermi level gives a finite mean free path at leading order in a random potential.
(We note that a regular metal is different from a Dirac-Weyl system even in the hypothetical limit of a vanishing Fermi energy since there is an energy gap between conduction and valence bands for the regular metal whereas a Dirac-Weyl system is gapless-- i.e. a regular metal simply becomes an ordinary gapped semiconductor in the zero Fermi energy limit whereas the gapless  Dirac-Weyl system is a nontrivial semimetal for zero Fermi energy.)
For the Dirac problem of interest here, from a scaling analysis of the action it is straightforward to
see the perturbative irrelevance in three dimensions of disorder for massless Dirac and Weyl fermions~\cite{Fradkin-1986}.
Thus the semimetal (SM) phase
 could be stable up to some non-zero critical disorder strength,
with a disorder-driven itinerant quantum critical point (QCP) into a so-called diffusive metal (DM) phase at higher disorder.
(We mention here for completeness that in two-dimensions, e.g. graphene, disorder is perturbatively relevant, and thus infinitesimal disorder
immediately converts undoped graphene from being a semimetal in the clean limit into a diffusive  metal ---
thus two is the perturbative lower critical dimensionality for the disordered Dirac-Weyl problem.)  The natural question that now arises with respect to the disordered three-dimensional Dirac-Weyl systems is whether the perturbative robustness of the semimetallic phase to disorder applies generally or is simply a perturbative result (perhaps to all orders in the perturbation theory), not valid in the nonperturbative theory.  The goal of the current work is to settle this question definitively.  Although the disordered Dirac-Weyl systems have been theoretically studied very extensively in the literature~\cite{Fradkin-1986,Goswami-2011,Kobayashi-2014,Brouwer-2014,Bitan-2014,*Bitan-2016,Nandkishore-2014,Pixley-2015,Altland-2015,Sergey-2015,Leo-2015,Sbierski-2015,Pixley2015disorder,Garttner-2015,Liu-2015,Bera-2015,Shapourian-2015,Altland2-2015,Sergey2-2015}, essentially all of this work, except for a very recent one (Ref.~\onlinecite{Nandkishore-2014}), study the properties of the disorder-driven SM-DM quantum phase transition, taking it for granted that such a disorder-induced QCP indeed exists in three dimensions following the predictions of the perturbative field theory.  Our current work reconciles the huge body of  QCP theoretical work in the literature with the
existence of nonperturbative or rare-region effects which 
lead to the `suppression' or `avoidance' of such a SM-DM QCP.

It is known that non-perturbative effects of rare
regions may give rise to a non-zero (albeit exponentially small) density of states at zero energy for
an infinitesimal strength of disorder~\cite{Nandkishore-2014,Wegner-1981}, thus converting the ballistic excitations in a weakly disordered SM to diffusive in the low energy limit,
which  thus results in the absence of a strict SM phase (with vanishing zero energy DOS).
It is not uncommon for
disorder to fundamentally change the nature of clean critical points (e.g. the Harris criterion~\cite{Harris-1974} says this happens
when the clean correlation length exponent $\nu < 2/d$ ), while the Chayes-Chayes-Fisher-Spencer (CCFS)~\cite{CCFS-1986} inequality
for the exact correlation length exponent of the disordered system
($\nu \geq 2/d$) applies to critical points that occur in the presence of quenched randomness.
Interestingly, the one loop perturbative renormalization group (RG) calculation of the critical exponents for the proposed SM to DM QCP are consistent with the CCFS inequality (since $\nu = 1$,  Refs.~\onlinecite{Fradkin-1986,Goswami-2011}) as, in fact, are the 2-loop RG calculations~\cite{Bitan-2014,*Bitan-2016,Sergey2-2015} and all numerical estimates in the literature~\cite{Kobayashi-2014,Brouwer-2014,Sbierski-2015,Pixley2015disorder,Liu-2015,Bera-2015}, and therefore it is not {\it a priori} obvious that rare region effects should change the universality of this transition.
Given the field theoretic RG analyses and the large body of direct numerical studies of the disorder-driven SM-DM QCP finding the various critical exponents and identifying the critical coupling as well as the apparent consistency between the theoretical (and numerical) correlation exponent with the CCFS inequality, it seems reasonable to assume that the rare regions arising out of nonperturbative disorder effects do not change the nature of the QCP in any substantial manner.
In this work we explore the fate of this SM-DM QCP using specialized numerical techniques that allow for the direct study of these rare region effects.  Our work definitively establishes that the putative SM-DM QCP becomes avoided or hidden due to the rare region effects, although the crossover effects of the avoided QCP show up in the numerical results.   This is thus consistent with all the earlier numerical work finding an apparent existence of the QCP, which, we now argue, strictly speaking does not exist when examined to the lowest energies.  Our work leads to the important conclusion that there is no disorder-driven SM-DM QCP in three-dimensional Dirac-Weyl systems,  only an avoided QCP, and the Dirac point develops nonzero (albeit exponentially small) DOS even for weak disorder.

\begin{figure}[t]
\centering
  \includegraphics[width=0.7\linewidth,angle=-90]{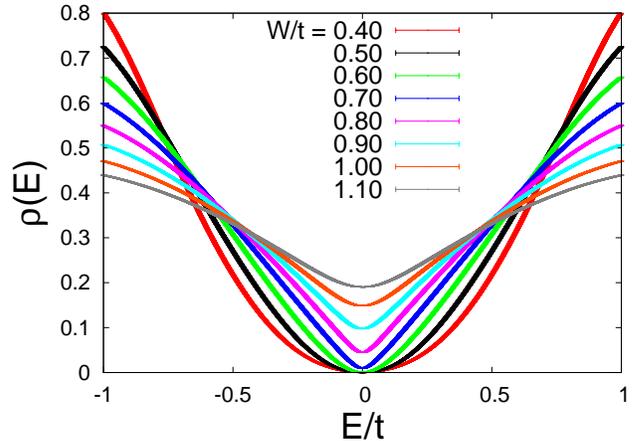}
\caption{(color online) The density of states (DOS) $\rho(E)$ versus energy $E$ for systems of linear size $L=71$ at KPM expansion order $N_C=2048$ averaged over twisted boundary conditions.
$1,000$ disorder realizations were used for each value $W$ of the disorder.  As $W$ approaches the avoided transition at $W \approx 0.75t$
 the DOS sharpens up, approaching $\rho(E)\sim |E|$ over some range of $|E|$.
Although $\rho(0)$ is nonzero at all values of $W$, this only becomes apparent on this linear scale for $W > 0.6 t$ where Ref.~\onlinecite{Pixley2015disorder} estimated the location of the QCP to be.}
\label{fig:0}
\end{figure}

To put the problem into context, we first review the existing evidence for the disorder-driven SM to DM QCP in undoped Dirac-Weyl systems.
The seminal work of Fradkin~\cite{Fradkin-1986} established the existence of this disorder driven (perturbatively accessible) QCP.  More recently, the properties of this proposed QCP have been calculated in a renormalization group treatment of the problem~\cite{Goswami-2011}, which has now been extended to two loops~\cite{Bitan-2014,*Bitan-2016, Sergey2-2015}.
 The field theory of the QCP can be constructed in terms of an interacting ``$Q^4$ theory'' (similar to a $\phi^4$ theory for magnetism but now $Q$ is the replicated matrix field)  strongly coupled to massless Dirac fermions~\cite{Pixley2015disorder}, while for the Weyl case due to topological considerations a separate field theory has been derived~\cite{Altland-2015,Altland2-2015}.
Tuning the clean model away from the Dirac/Weyl limit by varying the power law of the dispersion relation~\cite{Sergey-2015,Leo-2015}, this transition has been shown to occur
even in some one-dimensional models~\cite{Garttner-2015} (akin to a long-range Ising model). Thus, the existence of the putative SM-DM QCP seems to be well-established from a field theoretic perspective.

Due to the non-interacting nature of the problem, various numerical techniques which can reach rather large system sizes, have been used to study the properties of this QCP.  A main focus has been the direct calculation of the low energy density of states (DOS) $\rho(E)$, since the DOS is expected to be singular at $E=0$ at the transition.
Moreover, the DOS can be related to the critical exponents via the scaling hypothesis~\cite{Kobayashi-2014}.  Following this, the dynamic exponent $z$ and the correlation length exponent $\nu$ have now been numerically estimated for several models using the directly numerically calculated DOS~\cite{Kobayashi-2014,Pixley2015disorder,Garttner-2015,Liu-2015,Bera-2015}.  In addition to the DOS, the conductivity has also been studied across this transition~\cite{Brouwer-2014,Sergey-2015,Liu-2015,Shapourian-2015} and has led to estimates of $z$ and $\nu$ for a single Weyl cone~\cite{Sbierski-2015}, which are consistent with the exponents obtained from DOS calculations. In all of these numerical calculations the CCFS inequality $\nu \ge 2/d$ is well-satisfied.  It is important to mention that totally independent from this SM to DM transition, at a much larger disorder strength the Anderson localization transition has been established in some of these models~\cite{Pixley-2015, Liu-2015}. The current work is entirely in the low-disorder regime (where the SM-DM avoided QCP resides) and has nothing to do with the high-disorder Anderson localization transition from a DM phase to an Anderson insulator phase~\cite{Pixley-2015}, which occurs at roughly $W_l/t =  3.75$ for the model under consideration with Gaussian disorder (see the Appendix).

Despite all of this evidence for a stable SM phase and a SM-DM QCP in the presence of disorder, the effects of rare regions call their existence into doubt (and also raise the important and relevant question of why the extensive previous numerical work on the problem always indicates the
existence of such a SM-DM QCP).
As shown in Ref.~\onlinecite{Nandkishore-2014} through a Lifshitz tail~\cite{Lifshitz-1964} type analysis for the DOS~\cite{Yaida-2012},
rare quasi-localized eigenstates (``rare regions'') will contribute an exponentially small (in disorder strength) DOS at zero energy,
thus making the lowest energy excitations diffusive for arbitrarily weak disorder.
Therefore, in the strictest sense, there cannot be a disorder-driven SM-DM QCP in this problem since ``both'' phases must have nonzero DOS at zero energy although the rare region-induced DOS, being exponentially small, may very well be extremely difficult to discern (or more precisely, the SM-DM transition cannot
 have the DOS being zero in one phase and nonzero in the other phase as one of its features).
These rare eigenstates in the Dirac-Weyl case are distinct from traditional Lifshitz tail states in a band gap (e.g. of a regular semiconductor), as they are only quasi-localized (in contrast to the exponential nature of the disorder-induced Lifshitz band tail states in the semiconductor band gap), with the eigenfunctions falling off at short distances as a power law $\sim 1/r^2$ of the distance $r$ from the local extremes of the random potential,  and presumably being extended and weakly diffusive at much longer length scales.  However, none of the previous numerical studies on the SM-DM QCP has ever observed any signs of these ``elusive'' rare eigenstates, apart from possibly a large conductance tail in the data of Ref.~\onlinecite{Sbierski-2015}, whose relation to rare events has not been made clear (and which may very well be a finite size effect because finite size systems always have finite conductance).  Overall, the numerical data in recent papers seem consistent with the existence of a disorder-driven SM-DM QCP, with a notable agreement between analytical and numerical calculations of the dynamic exponent $z(\approx1.5$ within error bars).

In this paper, focusing on a particular lattice model of Dirac (and time-reversal symmetric Weyl) fermions in the presence of short range potential
disorder, we address these issues (i.e. both the QCP and rare regions on the same footing) by first finding the rare eigenstates
in the SM regime, and then
exploring the behavior of the model in the vicinity of the SM-DM avoided QCP.
We choose a relatively simple model that has been shown~\cite{Pixley-2015,Pixley2015disorder} to exhibit a sharp SM to DM transition (or crossover),
without the additional complications of mass terms.  Using (separately) Lanczos~\cite{Lanczos-1950,Lehoucq-1998} and the kernel polynomial method (KPM)\cite{Weisse-2006}
we provide definitive numerical evidence for the existence of two distinct types of low-$|E|$ eigenstates in the three-dimensional undoped (i.e. Fermi level at the band touching point taken to be the energy zero) Dirac-Weyl systems for weak disorder strengths.
Focusing first on the distribution of the first few low-$|E|$ eigenstates, we show for weak disorder that the DOS is well
described by ``Dirac peaks'' (the clean eigenstates that have moved and broadened in energy due to disorder) and an orders of
magnitude smaller (i.e. rarer) ``background'' that fills in between these finite-size Dirac peaks giving a nonzero contribution to $\rho(E=0)$.
We are able to systematically establish that the eigenstates that make up the peaks are perturbatively dressed Dirac eigenstates and the smaller background DOS comes from quasi-localized rare eigenstates.  As we show, the peak eigenstates are well described by perturbation theory and are Dirac plane waves
weakly distorted by the random disorder potential;
they disperse linearly from $E=0$.  The rare eigenstates are quasi-localized (i.e. the wavefunctions fall off algebraically rather than exponentially) and thus weakly dispersive.
Our numerical results indicate that these rare eigenstates arising from disorder (with no clean system analogs) are power-law localized like
$\sim 1/r^{x}$ at short distances $r$ from the site/cluster with the largest disorder strength with the power law $x$ in the
range $1.5-2.0$,
in excellent agreement with the analytic prediction ($\sim 1/r^2$).~\cite{Nandkishore-2014}

We estimate the zero energy DOS from the background rare region contribution using separately Lanczos and the KPM, finding good agreement between the two methods.
Our reason for using two completely independent numerical methods in identifying and quantifying rare region contributions to the DOS is to ensure the accuracy and consistency of our results, given the significance of our findings.
Over a range of about four orders of magnitude in the DOS, it well satisfies the rare region form $\rho(E=0)\sim\exp(-a/W^2)$, where $W$ is the amplitude of the random potential.  As the disorder strength $W$ increases, eventually there is a crossover to the avoided quantum critical (AQC) regime, where there no longer is a clear separation of the eigenstates between dispersive Dirac states and quasi-localized rare resonances, as the magnitudes of the DOS contributions from the dressed Dirac states and the rare regions start overlapping.  In this crossover AQC regime, the DOS far enough away from $E=0$ does show a quantum critical form
$\rho(E) \sim |E|^{(d/z)-1}$  with a $z\cong 1.5$ (Ref.~\onlinecite{Pixley2015disorder}), but this scaling behavior is cut off at lower energies.
Thus, we conclude that for the model under consideration (and other models with similar \emph{symmetry} considerations) the SM-DM QCP is converted by nonperturbative effects into an \emph{avoided} QCP, although the crossover effects of the AQCP manifest themselves at nonzero energies in spite of the QCP itself being suppressed.
Our results, taken together with previous work, are
consistent with a QCP that is ``hidden'' by effects that are nonperturbative in the disorder.
But a quantum critical regime still exists over a range of nonzero energies, where the rare region correction to quantum critical scaling is small and the nonzero energy behavior of the avoided QCP can therefore manifest itself.  The actual size of this crossover region in the energy/disorder space depends crucially on the nonuniversal details of the problem.  In other words, the nonzero value of $\rho(E=0)$ due to the rare eigenstates cuts off the divergence of the correlation length at some length scale $\xi_{RR}$, thus for length scales $\xi < \xi_{RR}$, and over the corresponding energy scales, the model looks critical.  This is why the previous
numerical studies~\cite{Kobayashi-2014,Pixley-2015,Pixley2015disorder,Garttner-2015,Liu-2015,Bera-2015,Sbierski-2015} observed an ``apparent'' SM-DM QCP. Our work thus not only establishes the nonexistence of the disorder-driven SM-DM QCP at the Dirac point due to rare region effects, but also reconciles the  large body of existing numerical work finding the existence of such a QCP by showing that the QCP becomes avoided at some large enough length scale (i.e. the correlation length never diverges in the thermodynamic limit) whereas at length scales smaller than this rare region induced cut off length scale, the observed behavior is consistent with a QCP.

Fig.~\ref{fig:0} gives an overview of the behavior:  At weak disorder in the SM regime (e.g. $W=0.4t$, where $t$ is the usual nearest-neighbor
kinetic hopping amplitude as in Eq.~(\ref{eqn:ham})), the DOS is close to the expected $\rho(E)\sim E^2$.  But actually there is a very small
nonzero DOS at $E=0$ that can not be seen on this linear plot.  In the avoided quantum critical regime
near $W=0.75t$, over a significant range of $|E|$ the DOS is closer to the expected QC behavior of $\rho(E)\sim |E|$, although this singularity is always rounded out near $E=0$ due to the rare region induced contribution which cuts off the quantum criticality. We are able to quantify how rounded out the singularity is by fitting the low-energy DOS to an analytic form.  Then in the DM regime at even larger values of $W$ the nonzero $\rho(E=0)$ becomes large.

The numerical work presented here establishes three distinct aspects of the disordered Dirac spectra: (1) well below the
putative SM-DM transition in the weak disorder regime,
there is a nonzero DOS at zero energy; (2) this nonzero DOS arises from the rare regions and is not due to dispersive Dirac quasiparticles, and obeys the
expected Ôrare regionÕ phenomenology (power-law quasi-localized eigenstates, and exponentially small DOS); (3) this converts the phase
transition in to an avoided QCP, which still exhibits a quantum critical regime, but at the lowest energies and the longest length scales becomes
an apparently nonsingular crossover between the diffusive metal regime and the regime of a semimetal with these rare quasi-localized eigenstates.

\section{Model and Methods}
\label{sec:model}
We study the effect of potential disorder on massless three-dimensional Dirac fermions on a simple cubic lattice in the presence of twisted periodic boundary conditions.  We consider the following Dirac Hamiltonian (introduced in Refs.~\cite{Pixley-2015,Pixley2015disorder})
\begin{equation}
H_D= \sum_{{\bf r},{\mu}={x},{y},{z}}\left( \frac{1}{2}i t_{\mu}\psi_{{\bf r}}^{\dag}\alpha_{\mu} \psi_{{\bf r}+\hat{\mu}} + \mathrm{H.c}\right)+\sum_{{\bf r}} \tilde{V}({\bf r})\psi_{{\bf r}}^{\dag}\psi_{{\bf r}},
\label{eqn:ham}
\end{equation}
where $\psi_{{\bf r}}$ is a four-component Dirac spinor and the $\alpha_{\mu}$ are the Dirac operators. We work in the Dirac representation
\begin{equation}
\alpha_{\mu} = \left( \begin{array}{cc}
0 & \sigma_{\mu}  \\
\sigma_{\mu} & 0 \end{array} \right),
\end{equation}
where $\sigma_{\mu}$ denotes the Pauli operators.
The clean dispersion relation is $E_0(\bk) = \pm t \sqrt{\sum_{\mu} \sin(k_{\mu})^2}$ for $t_x=t_y=t_z=t$,
and the model has eight Dirac points at ${\bf k}_D = \{ (0,0,0)$,$(\pi,0,0),(0,\pi,0),(0,0,\pi), $ $(\pi,\pi,0), (\pi,0,\pi),$ $ (0,\pi,\pi), (\pi,\pi,\pi) \}$.  The model has both time reversal symmetry and a continuous axial symmetry~\cite{Pixley-2015,Pixley2015disorder}.

\begin{figure}[h]
\centering
  \includegraphics[width=0.7\linewidth,angle=-90]{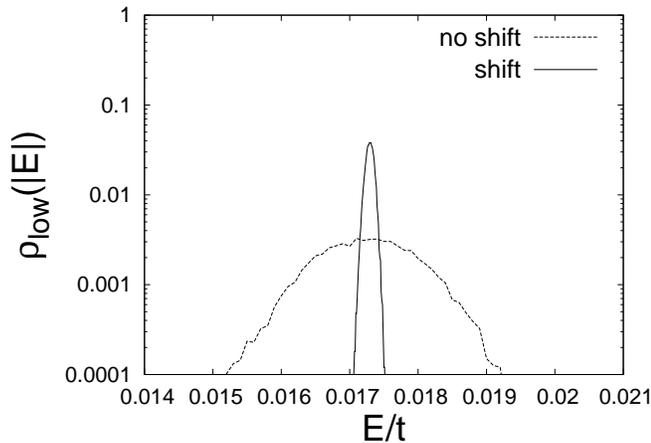}
\caption{The distribution of the absolute value of the lowest energy eigenvalue, i.e. $\rho_{\low}(|E|)$, in the absence and presence of a shift of the random potential, computed from Lanczos on $H^2$ for $L=55$, $W=0.3t$, $10,000$ disorder realizations, and a twist $\bm{\theta}=(\pi/3,0,0)$. Shifting the random potential dramatically sharpens up the width of the distribution of the lowest energy eigenvalue and has thus suppressed the leading finite-size effect.  The twist has put this peak at a nonzero energy, see Fig.~\ref{fig:dos_w=0.1}.}
\label{fig:shift_noshift}
\end{figure}

In order to get only one of the two degenerate eigenvalues associated with the conservation of axial charge we construct a two-component model defined as
\begin{equation}
H_W= \sum_{{\bf r},{\mu}={x},{y},{z}}\left( \frac{1}{2}i t_{\mu}\chi_{{\bf r}}^{\dag}\sigma_{\mu} \chi_{{\bf r}+\hat{\mu}} + \mathrm{H.c}\right)+\sum_{{\bf r}} \tilde{V}({\bf r})\chi_{{\bf r}}^{\dag}\chi_{{\bf r}},
\end{equation}
where $\chi_r$ is a two-component Pauli spinor, the $\sigma_{\mu}$ are the Pauli operators, and there is now only a degeneracy due to time reversal symmetry and the model represents a Weyl Hamiltonian. In the following we will only work with $H_W$ and from this point on refer to it as $H$.  We can remove time reversal symmetry by putting twisted boundary conditions on our samples of size $L\times L\times L$, so $t_{\mu} = t \exp(i \theta_{\mu}/L)$ with $-\pi <\theta_{\mu}<\pi$.  Then there are generally no degeneracies in a finite-size system, and the effect of the random potential at first order in perturbation theory is to simply rigidly move the energies of all these plane-wave eigenstates by the average value of the random potential, which is of order $L^{-3/2}$.  To remove this leading order finite-size effect, we shift the random potential to always have mean zero:
The unshifted random potential $V(\br)$ at each site is chosen independently from a Gaussian distribution with zero mean and standard deviation $W$.
We use $\tilde{V}(\br)$ to denote the shifted random potential with mean zero: $\tilde{V}({\bf r}) = V({\bf r})-V_0$ with
$\sum_{{\bf r}}V({\bf r})/L^3 = V_0$.  We always use energy units with $t=1$.

The clean system with $W=0$ has a spectrum that consists of discrete levels in any finite-size system.  Once we then average over the random potential at nonzero $W$ in the semimetal regime, these discrete Dirac energy levels each give a broadened peak in the DOS.  We want to minimize this broadening as much as possible in order to be able to see the rare quasi-localized eigenstates at low energies in between these Dirac peaks.  This is the motivation for shifting the random potential to always have zero average, see Fig.~\ref{fig:shift_noshift}.  This does not change the system at all in the limit of large $L$, but it changes the finite-size effects on the disorder-averaged DOS, making it easier to clearly see the rare region contributions in spite of their small values.

\begin{figure}[h]
\centering
\begin{minipage}{.5\textwidth}
  \centering
  \includegraphics[width=0.7\linewidth,angle=-90]{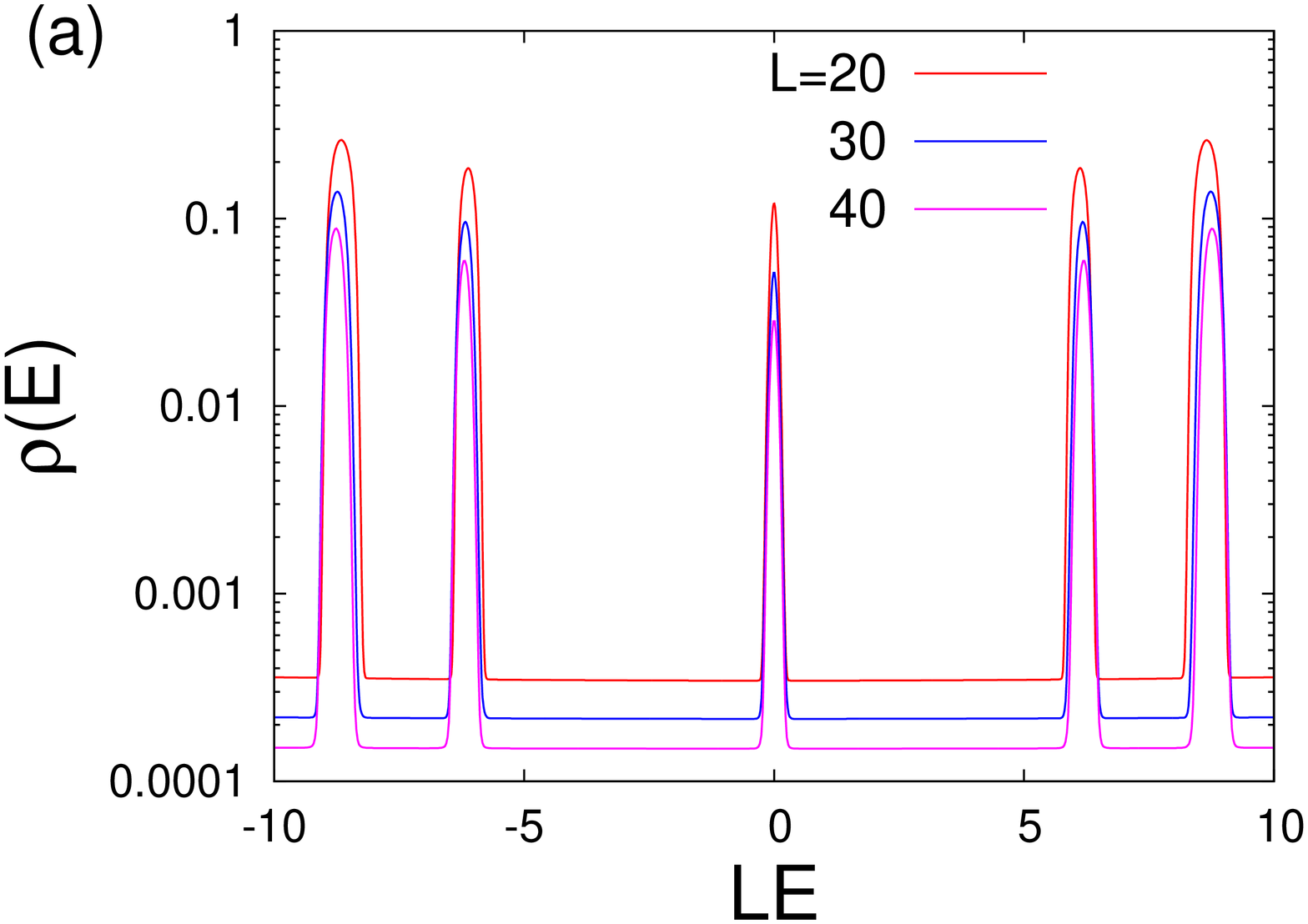}
\end{minipage}%
\newline
\begin{minipage}{.5\textwidth}
  \centering
  \includegraphics[width=0.7\linewidth,angle=-90]{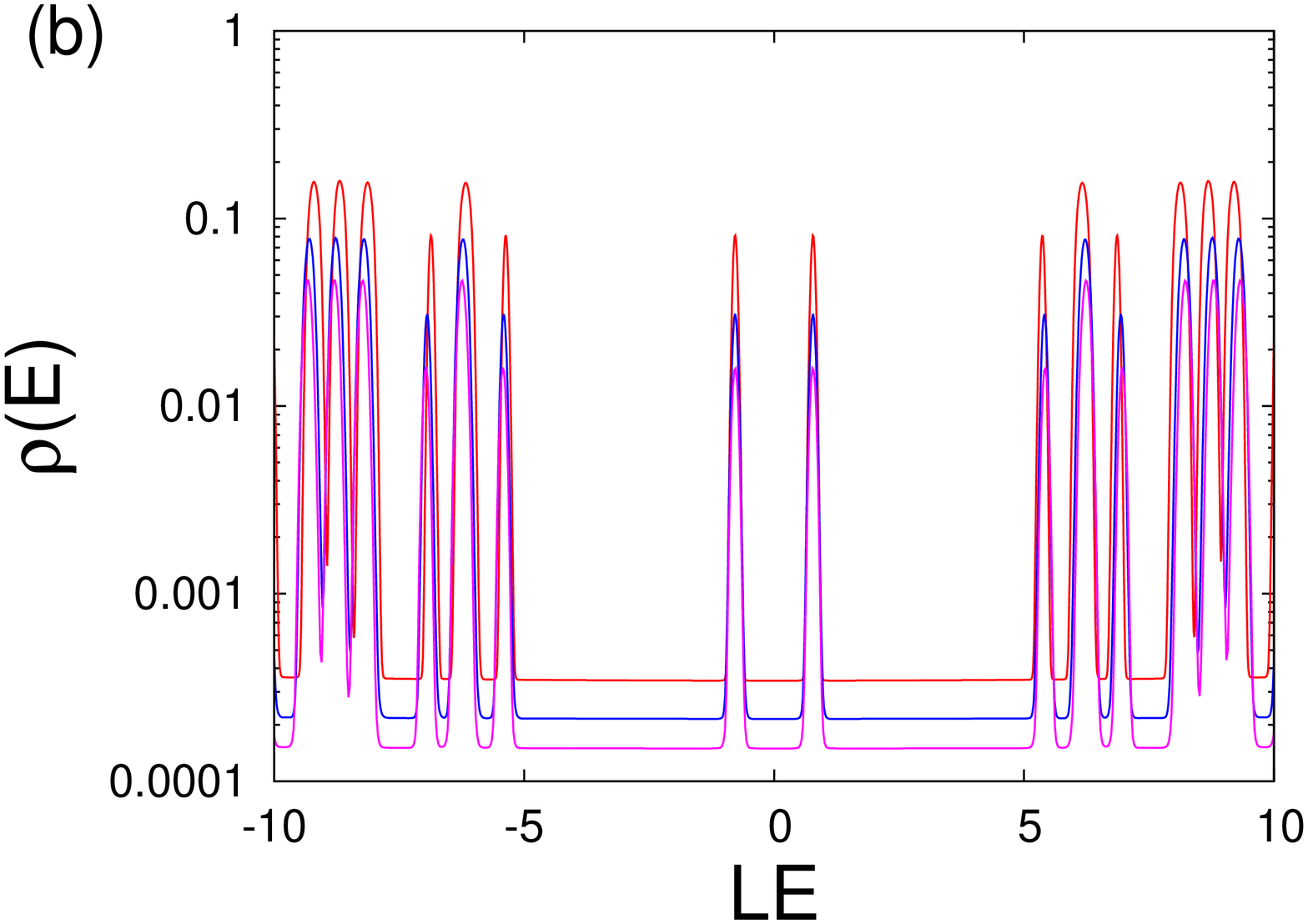}
\end{minipage}
\caption{(color online) Disorder-averaged density of states computed from the KPM for $W=0.1t$ as a function of $LE$ using the shifted random potential $\tilde{V}({\bf r})$ (a) without a twist and (b) with a twist of $\theta_x=\pi/4$.  The Dirac states that exist in the clean limit are broadened by the disorder, but remain well-separated.  Without a twist there is a Dirac peak at $E=0$.  Applying a twist splits this peak and pushes it away from zero energy, so that there are no states near $E=0$.  We have checked that in between the peaks there are no states and the flat background seen here is solely an artifact of the KPM.  We have also checked (not shown) that this KPM background is independent of $N_C$, provided $N_C$ is not too small. }
\label{fig:dos_w=0.1}
\end{figure}

One of the main results of this paper is to directly detect the
nonzero DOS at zero energy for weak disorder in the
semimetal regime, arising from rare quasi-localized eigenstates.  To do this, we use twisted boundary conditions such that in
the clean system ($W=0$) the DOS of a finite system does indeed strictly vanish at $E=0$.  Standard periodic boundary conditions for this system
unfortunately put Dirac states right at $E=0$, thus obscuring this question.
Introducing disorder broadens these disorder-averaged Dirac peaks, but for standard periodic boundary conditions the peak remains centered at
zero energy, as shown in Fig.~\ref{fig:dos_w=0.1}(a).  This connection of a clean Dirac state to its weakly disordered counterpart is made concrete in section~\ref{sec:eigenstates}, where we firmly establish that each eigenstate in the peak does represent a (perturbatively dressed) dispersive Dirac state.
We can push all of the Dirac states away from zero energy by using twisted boundary conditions (see Fig.~\ref{fig:dos_w=0.1}(b)),
which is achieved by $t_{\mu} = t \exp(i \theta_{\mu}/L)$ and using periodic boundary conditions.
For example, consider a twist of $\bm{\theta}=(\pi/4,0,0)$ as in Fig.~\ref{fig:dos_w=0.1}(b), this pushes the lowest energy eigenstates out to $E=\pm t |\sin(\pi/4L)|$ (for $W=0$), with no state closer to $E=0$.

For the Lanczos calculations that follow we consider odd $L$ and use a twist of $\bm{\theta}=(\theta_x,0,0)$, usually with $0<\theta_x<\pi/2$.
This is enough to lift the degeneracy of the eigenstates with the four lowest $|E|$.
Focusing on nondegenerate states is preferable for Lanczos, since it has difficulties with degenerate eigenvalues.  When we want to push all of the Dirac states as far away from zero energy as possible we use a twist of $\bm{\theta}=(\pi,\pi,\pi)$ with even $L$, which places the lowest energy eigenstates at $\pm t\sqrt{3} |\sin(\pi/L)|$; we find this to be helpful when we use the KPM to estimate the rare eigenstate contribution to the zero-energy DOS.  Finally, when we want to estimate $\rho(E)$ while minimizing the finite-size effects at all $E$, we use KPM and average over all possible twisted boundary conditions, as in Fig.~\ref{fig:0}.

\begin{figure}[h]
\centering
\begin{minipage}{.5\textwidth}
  \centering
  \includegraphics[width=0.7\linewidth,angle=-90]{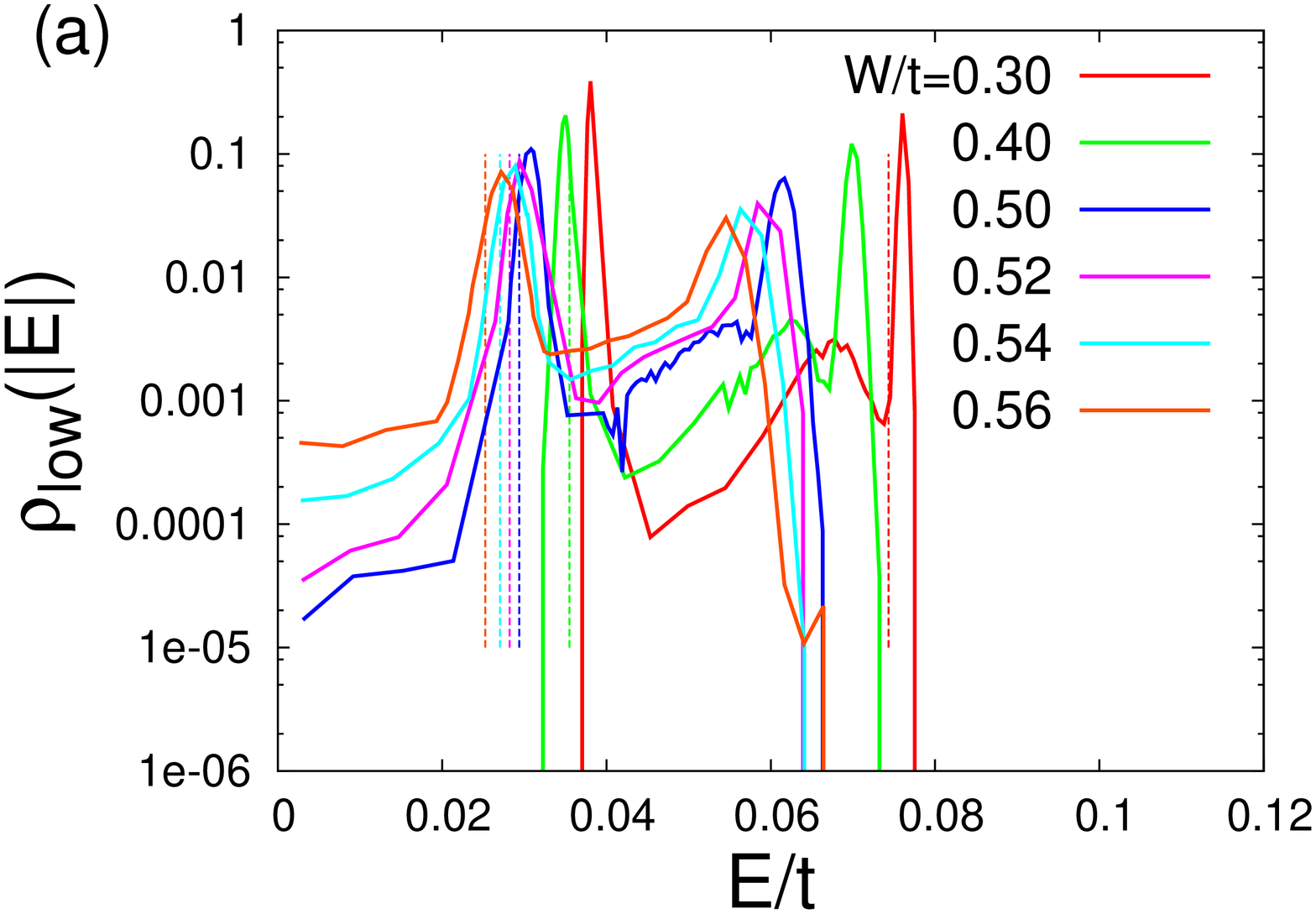}
\end{minipage}%
\newline
\begin{minipage}{.5\textwidth}
  \centering
  \includegraphics[width=0.7\linewidth,angle=-90]{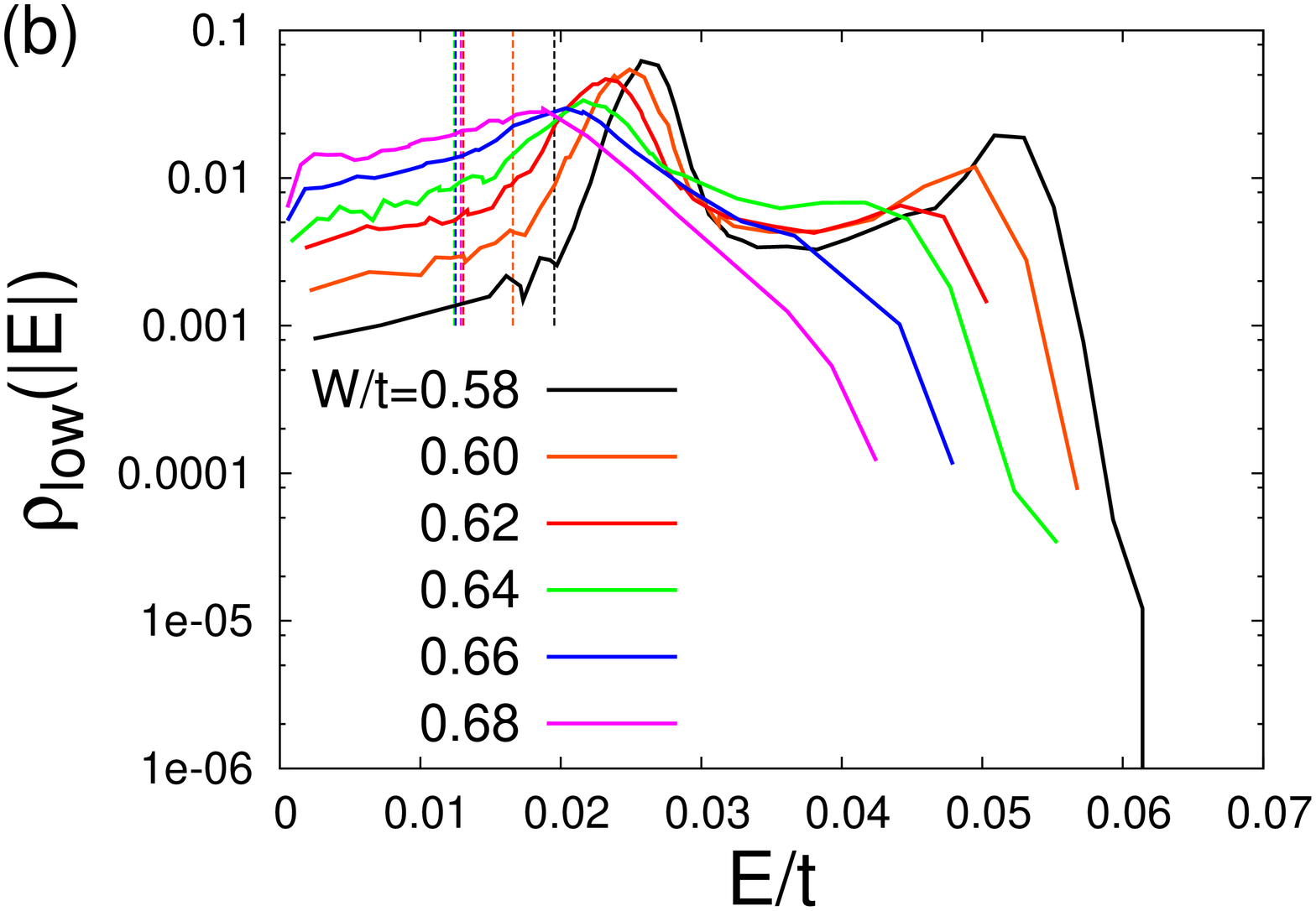}
\end{minipage}
\caption{(color online) Density of states computed from Lanczos on $H^2$ for the first four eigenstates, for system size $L=25$,
twist $\theta_x=\pi/3$ for (a) weak disorder and (b) for moderate disorder.  The vertical dashed lines mark $E_{N_u}^*(W,L)$
(see main text).  The two-component nature of the DOS, consisting of Dirac peaks and a smooth ``background'' is clear over an
intermediate range of disorder $W$ in the semimetal regime.  The background DOS extending down to zero energy is detected for $W\geq 0.5t$.
As $W$ increases, the Dirac peaks broaden and eventually the clear distinction between peaks and background is lost at larger $W$
in the avoided quantum critical regime.}
\label{fig:dos_L=25}
\end{figure}

We study the low energy eigenstates of $H$ using Lanczos and separately the KPM.
Lanczos provides accurate estimates of eigenstates and eigenenergies provided the spectrum has no near-degeneracies.
Therefore, we focus on the two-component model, usually with odd $L$ and a twist $\theta_x=\pi/3$.  We cannot use Lanczos at the largest disorder strengths, due to the spectrum becoming too dense near zero energy so Lanczos misses states due to the near-degeneracies.  In the semimetal regime, the four lowest-energy Dirac states for this twist and odd $L$ are at energies near $\pm E_0$ and $\pm 2E_0$, where $E_0\sim 1/L$ depends on both $L$ and $W$.
Even though disorder breaks the particle-hole symmetry $(E \rightarrow -E)$, for weak disorder it is only \emph{weakly} broken and Lanczos on $H^2$
combines states at the (approximately) same $|E|$ in to the same peak in $\rho(|E|)$, as shown in Fig.~\ref{fig:dos_L=25}.  In order to space out these
eigenvalues we also do Lanczos on $(H-(E_0(W,L)/4))^2$.
This puts the first four states near $|H-(E_0/4)|=3E_0/4, 5E_0/4, 7E_0/4, 9E_0/4$, effectively separating the peaks for each Dirac state, provided the width of each peak is not too broad.

Focusing on $N_u$ eigenstates from Lanczos the low energy average DOS can be computed from their distribution,
\begin{equation}
\rho_{\mathrm{low}}(|E|) = \frac{1}{N_R L^3} \sum_{r}^{N_R}\sum_{i=1}^{N_u} \delta (E - |E_i(r)|)
\end{equation}
for $N_R$ disorder realizations, where $E_i(r)$ is the $i$th eigenvalue of the $r$th disorder realization.  We often get the $N_u=4$ lowest states.  A few comments about the definition of $\rho_{\mathrm{low}}(|E|) $ are in order:  First, this definition implies that the DOS is normalized as the number of states per volume per $dE$, and therefore will have the same meaning as the full DOS for a particular energy $E$.  The (full) average DOS is defined as
\begin{equation}
\rho(E) = \frac{1}{N_R L^3} \sum_{r}^{N_R}\sum_{i=1}^{ D} \delta (E - E_i(r)).
\label{eq:dos}
\end{equation}
where $D=2 L^3$ is total number of states for the two-component model.
Second, the low energy DOS in Eq. (3) is only an accurate estimate of the full DOS for energies
$|E|\leq E_{N_u}^*$ where $ E_{N_u}^* = \min_{\{ r \}} |E_{N_u}(r)|$, i.e. the minimum value of the largest ($N_u$) eigenvalue Lanczos has
computed.  As a result, whenever we show $\rho_{\mathrm{low}}$ we will also plot vertical dashed lines to mark $E_{N_u}^*(W,L)$
(apart from Fig.~\ref{fig:shift_noshift} where this is not an issue due to the very weak disorder).
For $|E|>E_{N_u}^*$ Lanczos begins to miss some energy eigenvalues and this low energy estimate of the DOS will be depleted
relative to the full $\rho(E)$.  Lastly, since Lanczos has individual eigenvalue resolution (as opposed to the KPM),
we bin the results to generate a smoother estimate of the DOS.

For the KPM calculations presented in section~\ref{sec:dos} we consider twisted boundary conditions with $\bm{\theta} = (\pi,\pi,\pi)$. The technical details of the KPM can be found in Ref.~\onlinecite{Weisse-2006}.  KPM essentially trades off computing eigenvalues of $H$ for directly computing $\rho(E)$ via an expansion in terms of Chebyshev polynomials to an order $N_C$, and uses a kernel to filter out Gibbs' oscillations due to truncating the expansion.
(Avoiding direct diagonalization to calculate the eigenenergies allows the KPM to go to very large system sizes to directly compute the DOS, which would be inconceivable within an exact diagonalization.)
Here we use the Jackson kernel~\cite{{Weisse-2006}}, which amounts to replacing the delta function in Eq.~(\ref{eq:dos}) with a normalized Gaussian with a standard deviation $\sigma = \pi/N_C$, (in units of the bandwidth $\sim 2(\sqrt{3}t +W)$) so that the DOS over the full bandwidth remains normalized to $\int \rho(E) dE=$ (total number of states per volume).  In the calculations that follow we are using $N_C=1024$ unless otherwise stated.  When we average over samples, we use the same energy grid for all samples.  In order to effectively use the KPM to study the rare region contribution to the DOS, we find it essential to consider two issues:  First, due to this artificial broadening, using the twisted boundary conditions to push the Dirac peaks as far away from zero energy as possible is helpful, so that they do not contaminate the estimate of $\rho(E=0)$.  Second, even if there is a strictly zero DOS at a particular energy $E$, the KPM will give a non-zero number for $\rho(E)$ and will thus give an artificial ``background'' to the KPM DOS, for example see Fig.~\ref{fig:dos_w=0.1}.  As a result of this artificial background, even in the presence of the twist we find that the KPM  cannot accurately determine $\rho(0)$ for the smallest $W$ of interest.  Therefore, we can use the Lanczos to obtain $\rho(0)$ for weak disorder and the KPM for large disorder, whereas for intermediate disorder strengths we find that the estimates from the two methods do match consistently, which is an important numerical check for our results.

\section{Eigenstates of $H$}
\label{sec:eigenstates}
In this section we study the nature of the eigenstates of $H$.
As we discuss below, in the semimetal regime at weak disorder we find two qualitatively distinct types of eigenstates that give separate contributions to the DOS for finite samples.
In this regime, we find that the DOS can be separated into ``peaks'' and a ``background'' that lies in between the peaks.  This separation is useful as it will allow us to study the eigenstates that make up each contribution separately.  This is shown clearly in Fig.~\ref{fig:dos_L=25} for $L=25$ and 10,000 disorder realizations using Lanczos on $H^2$ for the first four lowest energy eigenstates.  Note that $H^2$ has put the Dirac peaks that are at the (approximately) same value of $|E|$ on top of each other, so in this figure the four states at $E\cong\pm E_0,\pm 2E_0$ produce two peaks.

For very weak disorder we see the two expected Dirac peaks with a ``background'' DOS developing between the peaks.  For larger disorder strengths in the semimetal regime the Dirac peaks remain and in addition we find the background is detected all the way down to zero energy.  We expect that the low energy tail is orders of magnitude too small for $W/t =0.30$ and $0.40$ to be observed for these system sizes and this number of disorder realizations, but is still actually present at any nonzero $W$.  We find this background DOS is an \emph{increasing} function of $|E|$.  For still larger disorder in the AQC regime the distinction between ``peaks'' and ``background'' is eventually lost.  For these larger disorder strengths we expect that the excitations are all diffusive and there are no longer well defined dispersive Dirac excitations; this occurs where the Dirac peaks are no longer visible $(W \cong 0.7t$).  For lower disorder strengths, the low energy background DOS represents quite rare eigenstates, e.g. for $W=0.5t$ the magnitude of the DOS at the peak versus the low energy (rare region) background is separated by almost four orders of magnitude. This explains why earlier numerical work invariably failed to find any rare region induced background DOS, thus concluding (erroneously) that the system remains a semimetal with zero DOS at $E=0$ up to the critical disorder strength.

Previous numerical studies of the SM-DM transition\cite{Kobayashi-2014,Pixley-2015,Pixley2015disorder,Garttner-2015,Liu-2015,Bera-2015} used periodic boundary conditions and even $L$, which produces a very strong finite-size effect on the zero-energy DOS in the semimetal.
In the current model with Dirac points  occurring at momenta commensurate with the lattice, even $L$ and periodic boundary conditions place the lowest energy eigenstates into the Dirac-peak centered around $E=0$.
In models where the Dirac/Weyl cone is located at momentum  incommensurate with the lattice, this is not the case. However, without carefully choosing $L$ or using a twist, Dirac/Weyl states will inevitably come sufficiently close to zero energy producing a strong finite size effect at $E=0$ (this is straightforward to see from our results since it is essentially equivalent to considering a small twist in our current model).
By pushing the Dirac states away from zero energy we have now allowed the background DOS at zero energy to be visible in the semimetal regime.  In the remainder of this section we will now study the two different types of eigenstates separately, i.e. the rare eigenstates that contribute to the low energy background DOS and the typical Dirac states that make up the peaks in the DOS.  In the next section we will estimate the background contribution to the zero energy DOS.

\begin{figure}[t!]
\centering
  \includegraphics[width=0.7\linewidth,angle=-90]{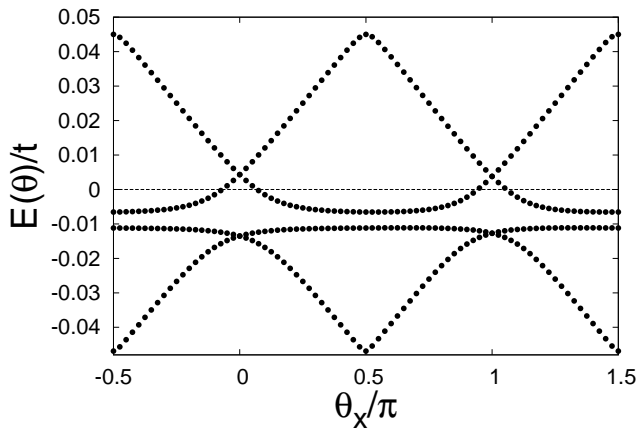}
\caption{The dispersion of the four lowest-$|E|$ eigenstates in the mini-zone for $L=25$ and $W=0.5t$ for a sample that shows two
quasi-localized eigenstates.}
\label{fig:disp_W=0.5}
\end{figure}
\subsection{Rare eigenstates}
In Ref.~\onlinecite{Nandkishore-2014} the theory of rare quasi-localized eigenstates in three-dimensional Dirac systems has been derived from a Lifshitz tail type formalism with the essential idea that the rare region effects in the Dirac-Weyl gapless systems are basically like the resonant versions of Lifshitz tail states which reside in energy band gaps.  It was found that the eigenstate that corresponds to a rare event, i.e. a disorder configuration that has either a site or a small group of sites that has a very large disorder strength, is a quasi-localized resonance that decays from the site $\br_{\mathrm{max}}$ with maximal disorder in a power law fashion like $\psi(r) \sim 1/r^2$ for $r/L \ll 1$ and $r \equiv |\br - \br_{\mathrm{max}}|$ more than one lattice spacing.  It is important to stress that the existence of these rare eigenstates of $H$ is non-perturbative in the disorder strength, and hence is outside the scope of the perturbative field theoretic analysis of the SM-DM QCP discussed in the literature.

We will now study the properties of eigenstates in a particular rare disorder realization that gives rise to
states in the background DOS.  In this subsection, we focus on a sample of size $L=25$ and a disorder strength $W=0.5t$.  By varying the twist $\theta_x$ in the $x$-direction we can determine the dispersion of an eigenstate in the ``mini'' Brillioun zone for momentum $-\pi/2 < \theta_x < 3\pi/2$.  Focusing on the lowest four eigenstates of $H^2$, we determine the sign of each eigenvalue of $H$ from its corresponding eigenvector, and construct the dispersion of these four eigenstates in both positive and negative energies as shown in Fig.~\ref{fig:disp_W=0.5}.  There are two weakly dispersive and thus quasi-localized states and two dispersive Dirac states, and these states hybridize near avoided level crossings.  The states come in pairs with opposite spin.  For twist $\theta_x=0$ and $\pi$ the system has time reversal symmetry and thus degenerate Kramers doublets.  In this sample, the rare states have a small negative energy, but among samples with such states, the energy is smoothly distributed through zero energy, resulting in a nonzero contribution to the zero-energy density of states.

\begin{figure}[t!]
  \centering
  \includegraphics[width=0.75\linewidth]{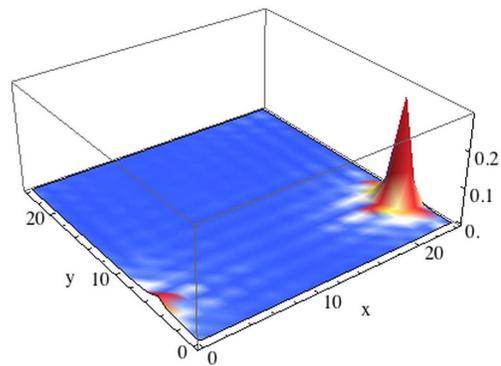}
\caption{(color online) Projected probability density $\sum_z |\psi(x,y,z)|^2$ versus $x$ and $y$ for a weakly-dispersing rare state with $L=25$, $W=0.5t$, and a twist $\theta_x=\pi/2$.  Note the system has periodic boundary conditions (with $t_x = t\exp(i \theta_x/L))$, and we have set the lattice spacing to unity. }
\label{fig:wf_W=0.5}
\end{figure}

Now that we have determined how this rare state disperses, we turn to the magnitude of the wavefunction $\psi(\br) \equiv \sqrt{|\psi_a(\br)|^2+|\psi_b(\br)|^2}$ where $a$ and $b$ label the two spinor components for the lowest-$|E|$ eigenstate with a twist that makes the state the most weakly dispersing, i.e. at $\theta_x = \pi/2$.  For plotting purposes we show $\sum_z |\psi(x,y,z)|^2$, which clearly shows a (quasi) localized wave-function, see Fig. 6.  We find the location in real space where two neighboring sites have a large disorder strength $V_i \sim 3 W$ at the same location $\br_{\mathrm{max}}$ where the wave-function's magnitude is maximal.  The probability of this disorder configuration is quite rare, relative to the probability of a typical configuration ($V_i \sim W$) it occurs with a probability $\sim \exp(-9)$, and therefore this is indeed a rare eigenstate.

\begin{figure}[t!]
\centering
\begin{minipage}{.5\textwidth}
  \centering
  \includegraphics[width=0.7\linewidth,angle=-90]{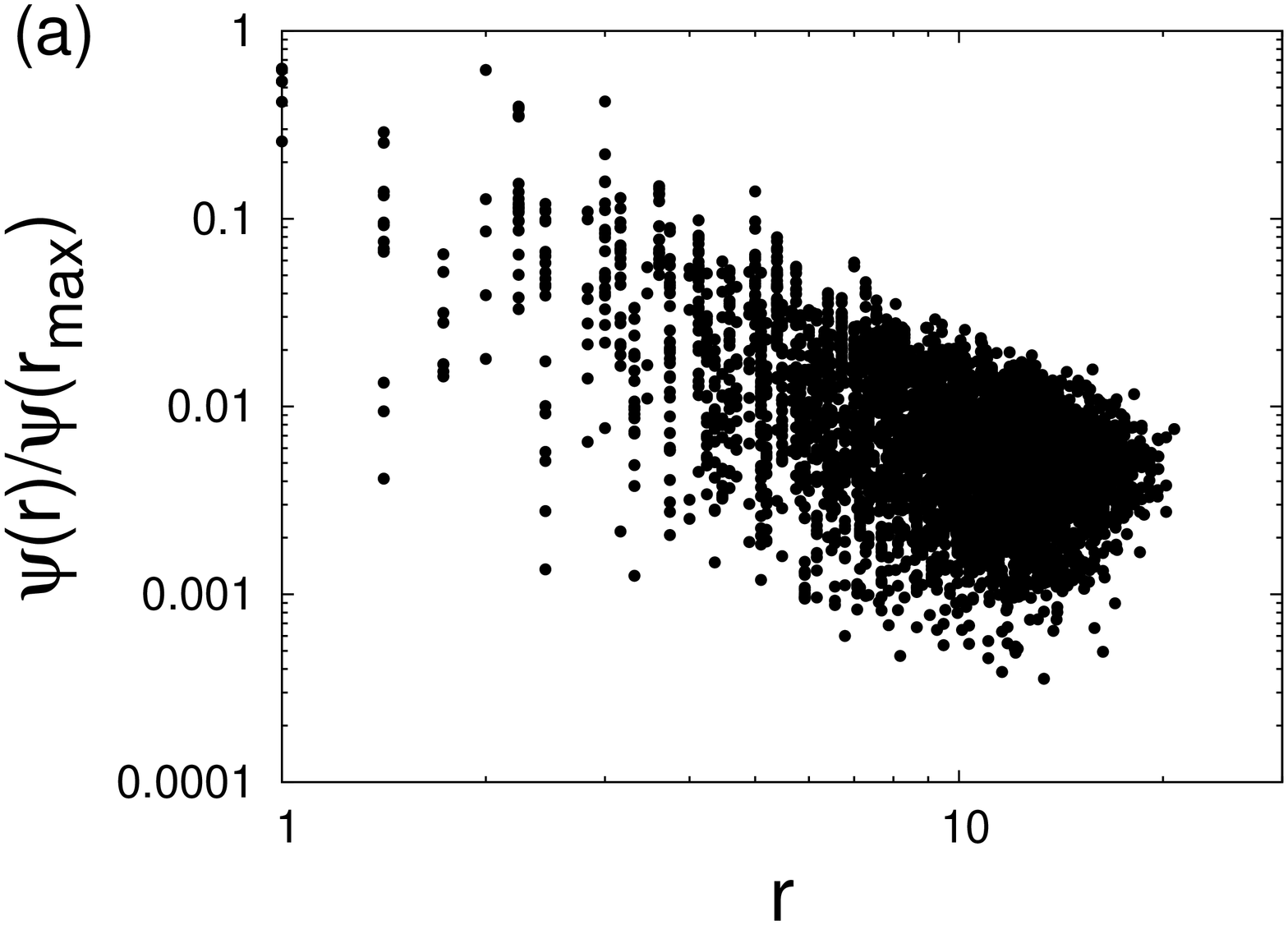}
\end{minipage}%
\newline
\begin{minipage}{.5\textwidth}
  \centering
  \includegraphics[width=0.7\linewidth,angle=-90]{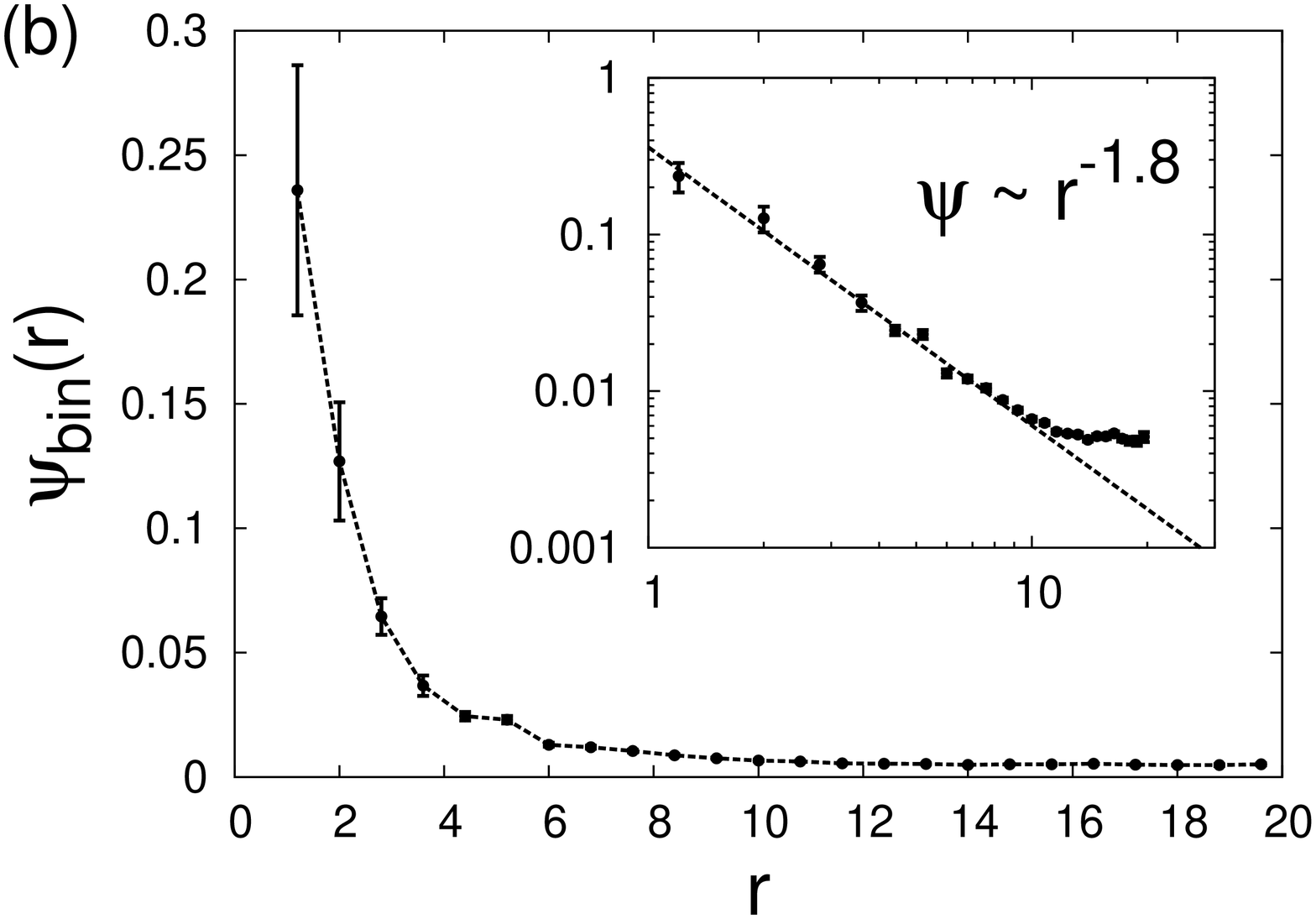}
\end{minipage}
\caption{Decay of the wave-function for a weakly-dispersing rare state from its maximal value with $L=25$, $W=0.5t$, and a twist $\theta_x=\pi/2$. (a) Scatter plot of the wavefunction as a function of $r$, the distance to the site $\br_{\mathrm{max}}$ with the maximal wavefunction value.  (b) Binned and averaged wavefunction $\psi_{\mathrm{bin}}(r)$ versus $r$, approximate error bars are obtained from the error on the mean of each bin under an assumption that the wavefunction magnitudes are uncorrelated from site to site.  (Inset) Log-log plot for  $\psi_{\mathrm{bin}}(r)$ versus $r$ displaying the power law decay $\sim 1/r^{1.8}$, the dashed line is the fit to a power law form.}
\label{fig:wf_decay}
\end{figure}

\begin{figure}[h!]
  \centering
  \includegraphics[width=0.7\linewidth, angle=-90]{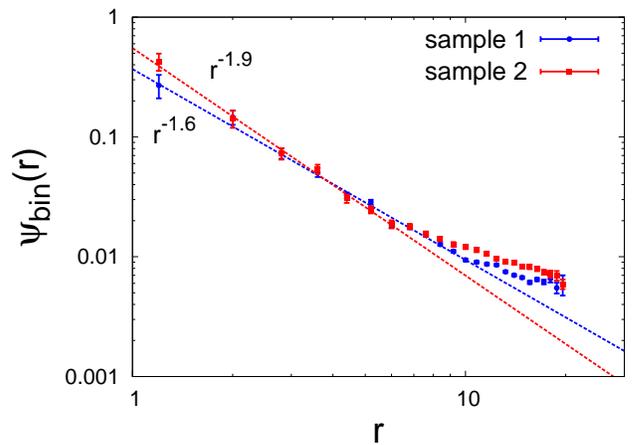}
  \caption{(color online)  Decay of the binned and averaged wave-function  from its maximal value for two different disorder samples that produce weakly-dispersing rare states with $L=25$, $W=0.6t$, and a twist $\theta_x=\pi/2$. Approximate error bars are obtained from the error on the mean of each bin under an assumption that the wavefunction magnitudes are uncorrelated from site to site.  The power law fluctuates sample to sample with a power law decay $\sim 1/r^{x}$, for $x=1.6$ (blue circles) and $1.9$ (red squares), the dashed line is the fit to a power law form.}
\label{fig:wf_decay2}
\end{figure}

We define the decay of the wave-function from its maximal value by computing $\psi(r) \equiv \psi(|\br - \br_{\mathrm{max}}|)$ (for $|r^{\mu} - r_{\mathrm{max}}^{\mu}|<L/2$ respecting the periodic boundary conditions). In Fig.~\ref{fig:wf_decay}(a) we show the scatter plot of the decay of the wavefunction from its maximal value, which indicates a power law trend in the data.
We then discretize the $r$ axis into bins and then average the value of $\psi(r)/\psi(\br_{\mathrm{max}})$ in each bin, which yields $\psi_{\mathrm{bin}}(r)$ as shown in Fig.~\ref{fig:wf_decay}(b) for $W=0.5t$.  To demonstrate the sample to sample variations we also show $\psi_{\mathrm{bin}}(r)$ for $W=0.6t$ in Fig.~\ref{fig:wf_decay2} for two different disorder realizations that give rise to distinct quasi localized eigenstates.  We now reach one of our main results, where over a range of $r$ we find the (binned) rare wave-function decays like
\begin{equation}
\psi_{R}(r) \sim \frac{1}{r^{x}},
\end{equation}
with a power law exponent $x$ that varies realization to realization of disorder in the range $1.5 - 2.0$, which is in excellent agreement (within numerical accuracy) with the analytic prediction of $\psi(r) \sim 1/r^{2}$ in Ref.~\onlinecite{Nandkishore-2014}.

It is interesting to contrast these eigenstates with states in the Lifshitz tail of the DOS in the presence of a band gap.  The latter are \emph{exponentially} localized~\cite{Anderson-1958} at a site with a very large disorder and contribute an exponential tail to the DOS near the band edge~\cite{Lifshitz-1964}.  Here, there is no band gap, and as a consequence these rare eigenstates are only power-law bound on these short length scales, and it is in this sense that they are only quasi-localized.  Intuitively, these rare states ``pull'' some spectral weight out of the Dirac bands and can place that weight at arbitrarily low energy, thus contributing to a background DOS that remains nonzero through $E=0$. Thus, the rare regions destroy the simple distinction between the SM phase (with zero DOS at zero energy) and the DM phase (with nonzero DOS at zero energy) as the DOS is always nonzero albeit very small for weak disorder.  The system is unstable to any disorder which immediately produces a nonzero DOS at zero energy.

\begin{figure}[h!]
  \centering
  \includegraphics[width=0.75\linewidth]{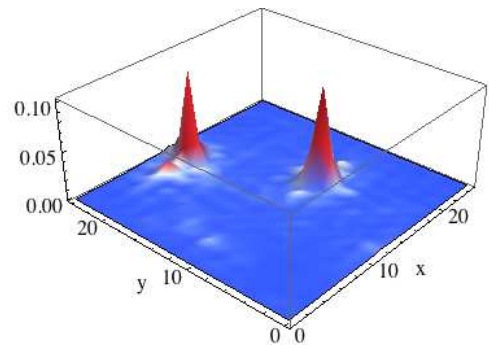}
\caption{(color online) Projected probability density $\sum_z |\psi(x,y,z)|^2$ versus $x$ and $y$ for a bi-quasi-localized rare state in the low energy tail of the DOS with $L=25$, $W=0.66t$, and a twist $\theta_x=0.325\pi$.  Note the system has periodic boundary conditions (with $t_x = t\exp(i \theta_x/L))$, and we have set the lattice spacing to unity.}
\label{fig:wf_W=0.66}
\end{figure}

For increasing disorder strengths that remain in the SM regime, the probability to generate more then one rare region increases, which makes it increasingly likely to find multiple quasi localized power law states in a single wavefunction per sample. Again focusing on an eigenstate that contributes to the low energy background DOS for $W=0.66t$, in Fig.~\ref{fig:wf_W=0.66} we show $\sum_z |\psi(x,y,z)|^2$ which clearly reveals the existence of a bi-quasi-localized wave function. Since each wavefunciton falls off (roughly) as $1/r^2$ from the sites ($r_1$ and $r_2$),
the overlap of these two quasi-localized peaks produces a non-zero tunneling matrix element that goes as $t_{RR}(r_1-r_2)\sim 1/|{\bf r}_1-{\bf r}_2|^2$.  Therefore it is natural to expect this tunneling will produce a diffusive metal where the conductance is mediated by hopping between these rare regions of large probability amplitude~\cite{Nandkishore-2014}.  We do expect that such wave functions are also generated at much weaker disorder, however their probability is so small that they are essentially never found in these size samples.

\begin{figure}[t!]
\centering
  \includegraphics[width=0.7\linewidth,angle=-90]{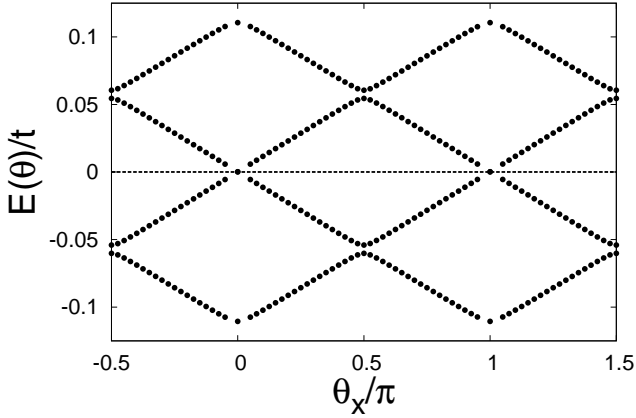}
\caption{The dispersion of four Dirac eigenstates in the mini-zone for $L=25$ and $W=0.3t$.}
 \label{fig:disp_W=0.3}
\end{figure}

\subsection{Perturbatively dressed Dirac eigenstates}

\begin{figure}[b!]
  \centering
  \includegraphics[width=0.75\linewidth]{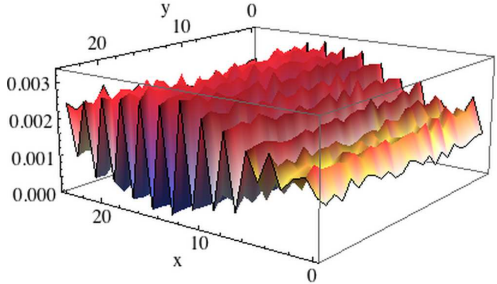}
\caption{(color online) Projected probability distribution $\sum_z |\psi(x,y,z)|^2$ for a linearly-dispersing Dirac eigenstate for $L=25$, $W=0.3t$, and a twist $\theta_x=\pi/2$.}
\label{fig:wf_W=0.3}
\end{figure}

We now consider the eigenstates that make up the low energy Dirac peaks in the DOS.  For $W=0.3t$ and $L=25$, by varying the twist we
again determine how the four lowest-$|E|$ states disperse in the mini-zone for a typical sample.  As shown in Fig.~\ref{fig:disp_W=0.3},
we find that the states disperse \emph{linearly,} just as they do in the absence of the random potential.  The only visible effects of
the random potential are a small renormalization of the Fermi velocity that we discuss below, and weakly-avoided level crossings at
$\theta_x=\pm\pi/2$, where the resulting eigenstates are standing waves.  The probability density in one of these standing wave eigenstates
is shown in Fig.~\ref{fig:wf_W=0.3}.  Here we can see that the eigenstate is very regular, with only weak ``noise'', due to the perturbatively irrelevant disorder.

\begin{figure}[t!]
\centering
\begin{minipage}{.5\textwidth}
  \centering
  \includegraphics[width=0.7\linewidth,angle=-90]{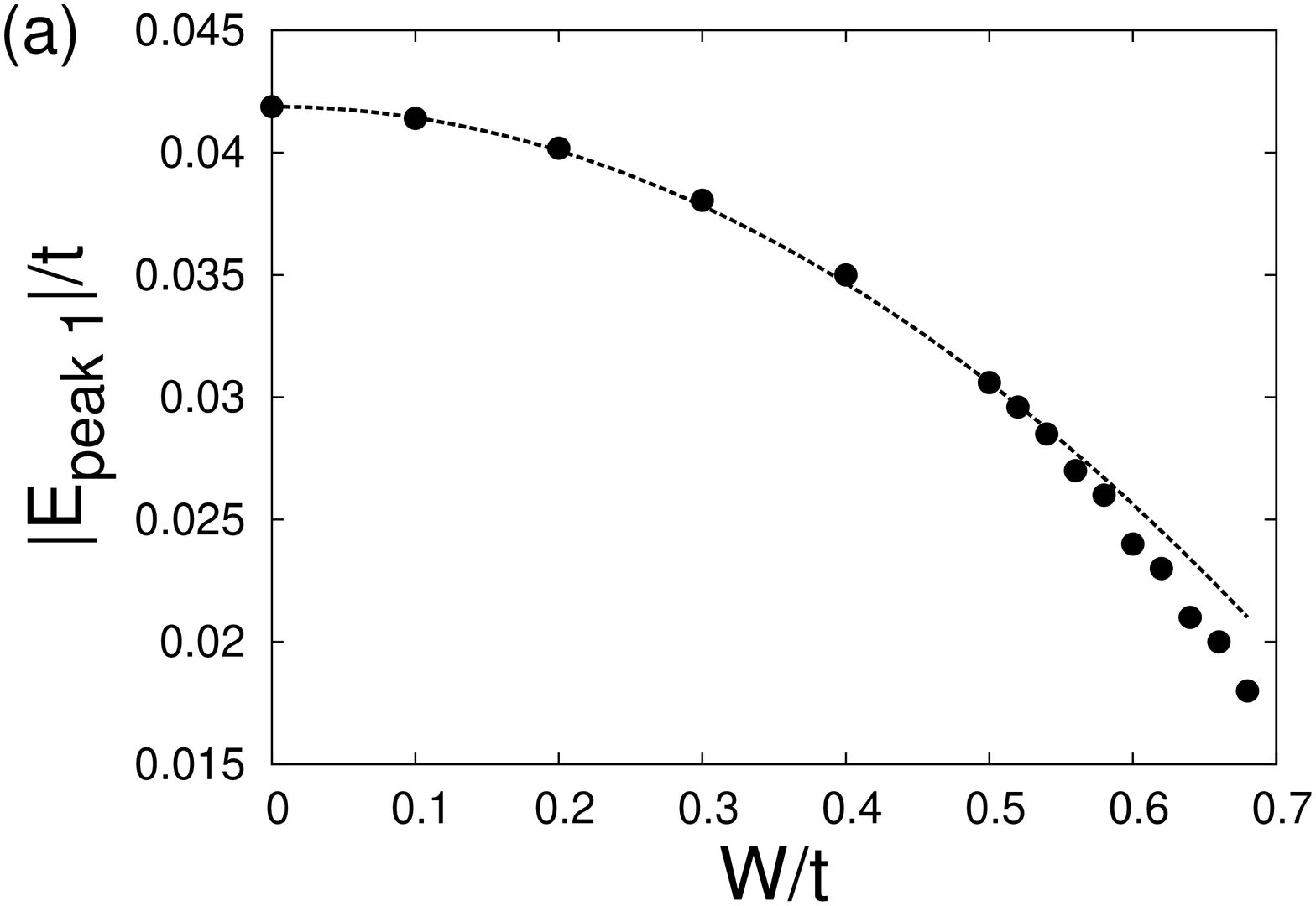}
\end{minipage}%
\newline
\begin{minipage}{.5\textwidth}
  \centering
  \includegraphics[width=0.7\linewidth,angle=-90]{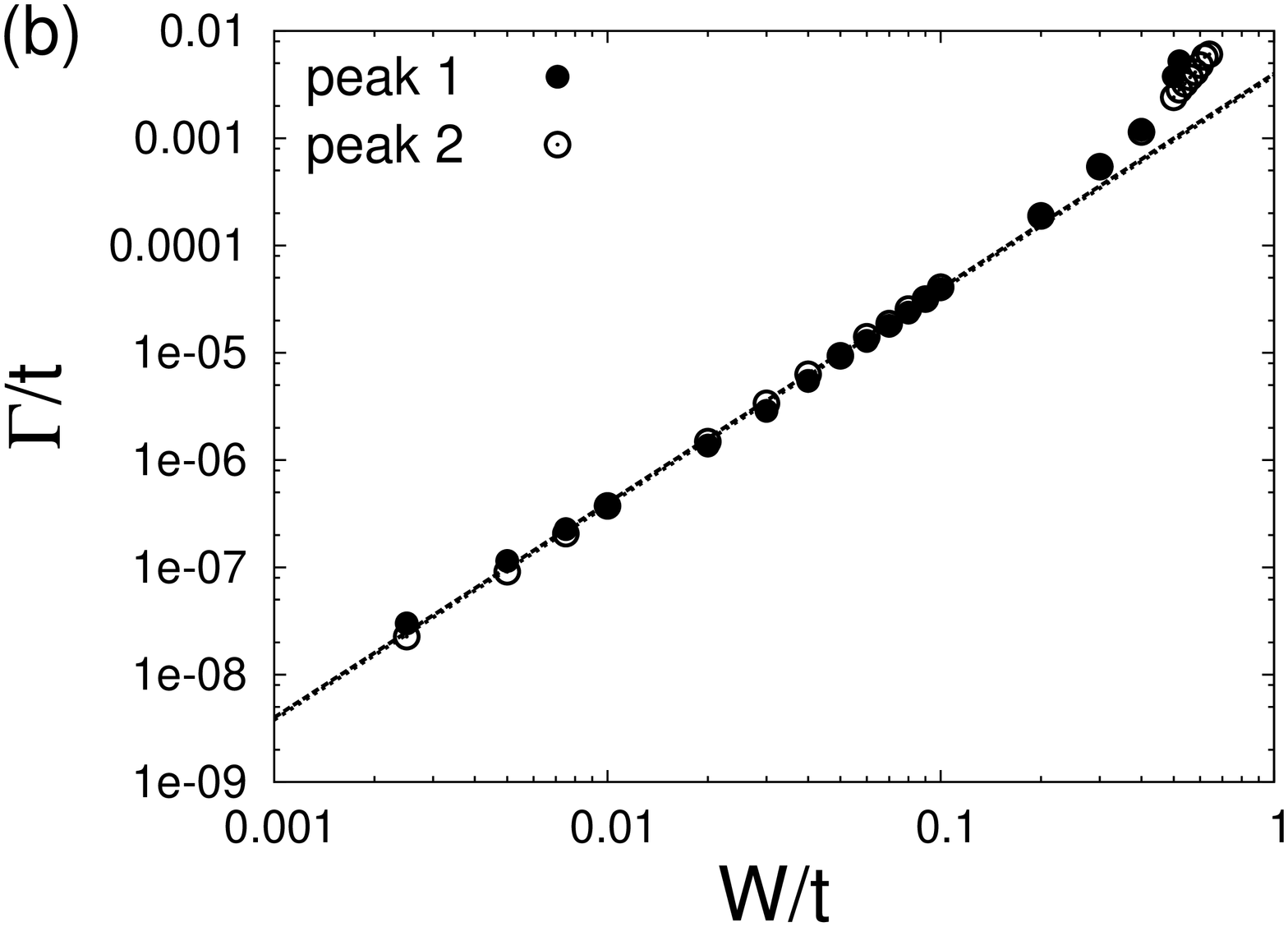}
\end{minipage}
\caption{Dependence of the Dirac peaks on $W$ with $L=25$ determined from Lanczos with a twist $\theta_x = \pi/3$.  (a) The energy of the first Dirac peak from $H^2$ in Fig.~\ref{fig:dos_L=25} versus $W$.  The dashed line is a fit to $E_1(W) = E_{1}(0)-aW^2$ with $a$ the only fit parameter: we see the fit works reasonably well up until about $W\approx 0.55 t$.  (b) Dependence of the FWHM $\Gamma$ of the first two peaks of $(H-E_0/4)^2$ (splitting the first peak in Fig.~\ref{fig:dos_L=25} into two) as a function of the disorder strength $W$, the dashed line is a fit to the perturbative form $b W^2$ with fit parameter $b$.  We find the fit works reasonably well for $W/t \le 0.1$ and this FWHM increases more strongly than quadratically in $W$ for larger disorder strengths. }
\label{fig:E_vs_W}
\end{figure}

We now focus on the quantitative properties of the lowest-$|E|$ Dirac peak at twist $\theta_x=\pi/3$, and their dependence on $W$ and $L$ (see Appendix~\ref{sec:appendix-A} for the perturbative analysis which is consistent with our numerical results for states in the Dirac peaks).  By restricting the random potential to have sum zero, we have eliminated the first-order-in-$W$ perturbative effect.  At order $W^2$, there is level repulsion from all other momenta, which is dominated
by the many states that are far away in energy, since the DOS is so small at low $|E|$.  The net effect of all this level repulsion is to reduce the Fermi velocity at order $W^2$, because positive (negative) energy states have stronger repulsion from the other positive (negative) energy states, since they are closer in energy, and thus the mean energy is pushed down (up) by the level repulsion.  This is illustrated in Fig.~\ref{fig:E_vs_W}(a), where we see a $\sim W^2$ suppression of the energy fits well over most of the semimetal regime.  The random component of the level repulsion gives the sample-averaged Dirac peaks a linewidth  $\sim W^2/L^{2}$ for small W (see Fig.~\ref{fig:E_vs_W}(b) for the full-width-at-half-maximum (FWHM) $\Gamma$ versus $W$).  Here we find the dependence on $W$ fits this quadratic behavior only for quite small $W$, and the linewidth increases faster than $\sim W^2$ throughout most of the SM regime.

\begin{figure}[t!]
\centering
\begin{minipage}{.5\textwidth}
  \centering
  \includegraphics[width=0.7\linewidth,angle=-90]{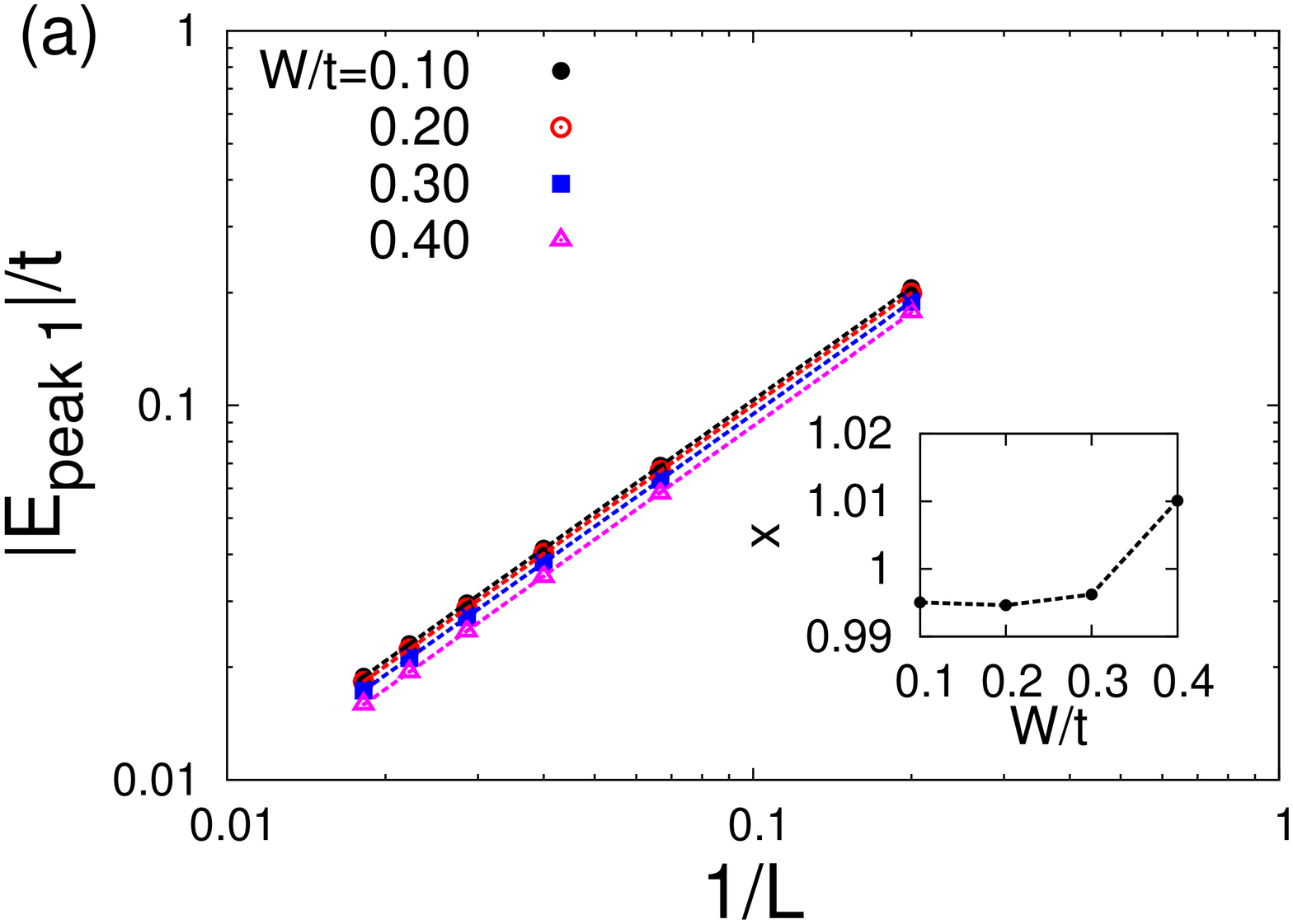}
\end{minipage}%
\newline
\begin{minipage}{.5\textwidth}
  \centering
  \includegraphics[width=0.7\linewidth,angle=-90]{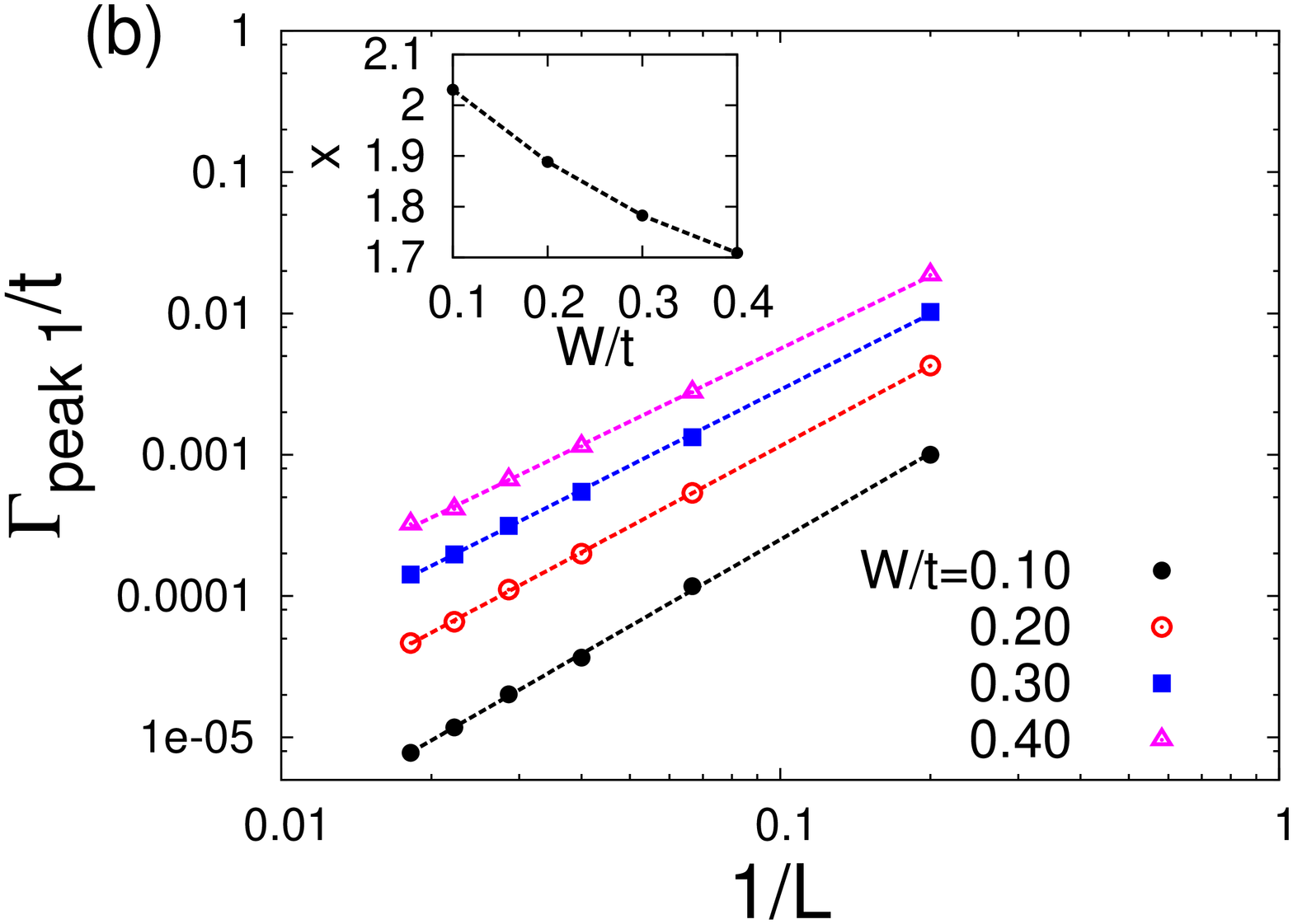}
\end{minipage}
\caption{(color online) Dependence of the energy and width of the first Dirac peak as a function of $L$ for various $W$, determined from Lanczos on $H^2$ with a twist $\theta_x=\pi/3$. The dashed lines are fits to the power law form $1/L^x$ and the values of $x$ versus $W$ are shown in each inset.  (a) The energy of the first Dirac peak versus $1/L$ and (b) the width of the first Dirac peak versus $1/L$.  We find the expected perturbative power law dependence ($\sim 1/L$)  for the energy, but the FWHM begins deviating from the pertubative result ($\sim 1/L^2$) for $W>0.1t$ consistent with Fig.~\ref{fig:E_vs_W}(b).}
\label{fig:E_vs_L}
\end{figure}

\begin{figure*}[t!]
\centering
\begin{minipage}{.33\textwidth}
  \centering
  \includegraphics[width=0.7\linewidth,angle=-90]{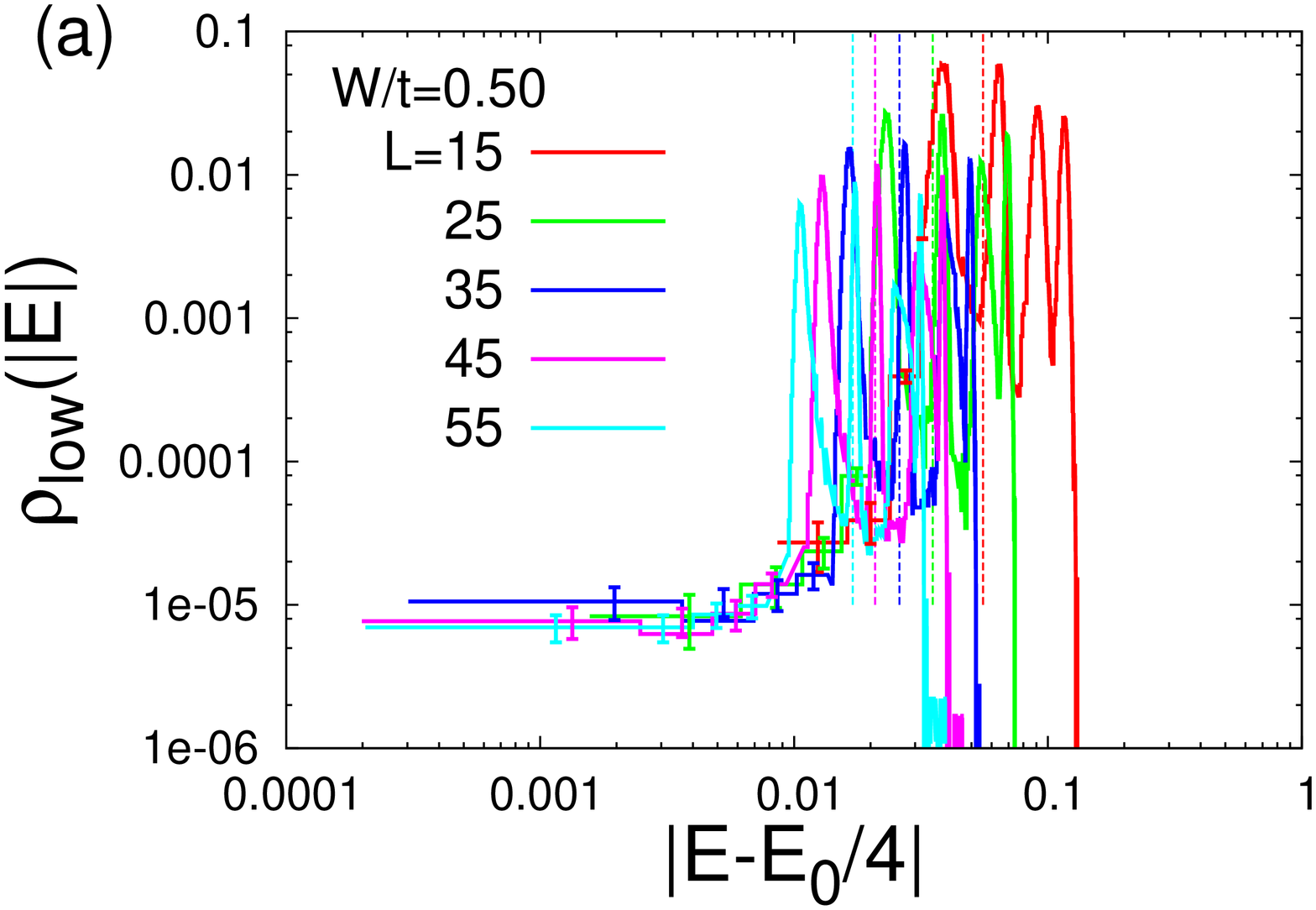}
\end{minipage}%
\begin{minipage}{.33\textwidth}
  \centering
  \includegraphics[width=0.7\linewidth,angle=-90]{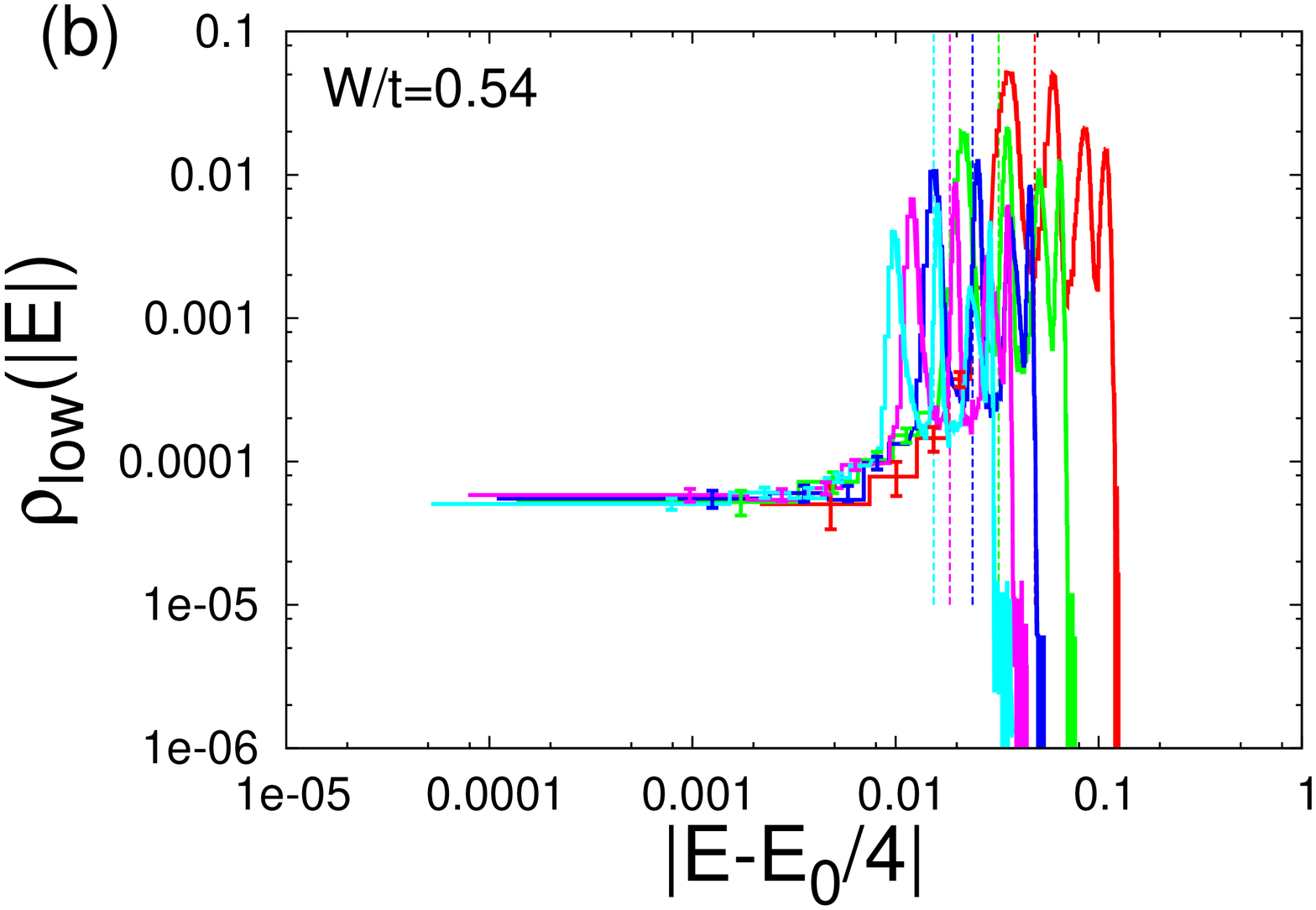}
\end{minipage}
\begin{minipage}{.33\textwidth}
  \centering
  \includegraphics[width=0.7\linewidth,angle=-90]{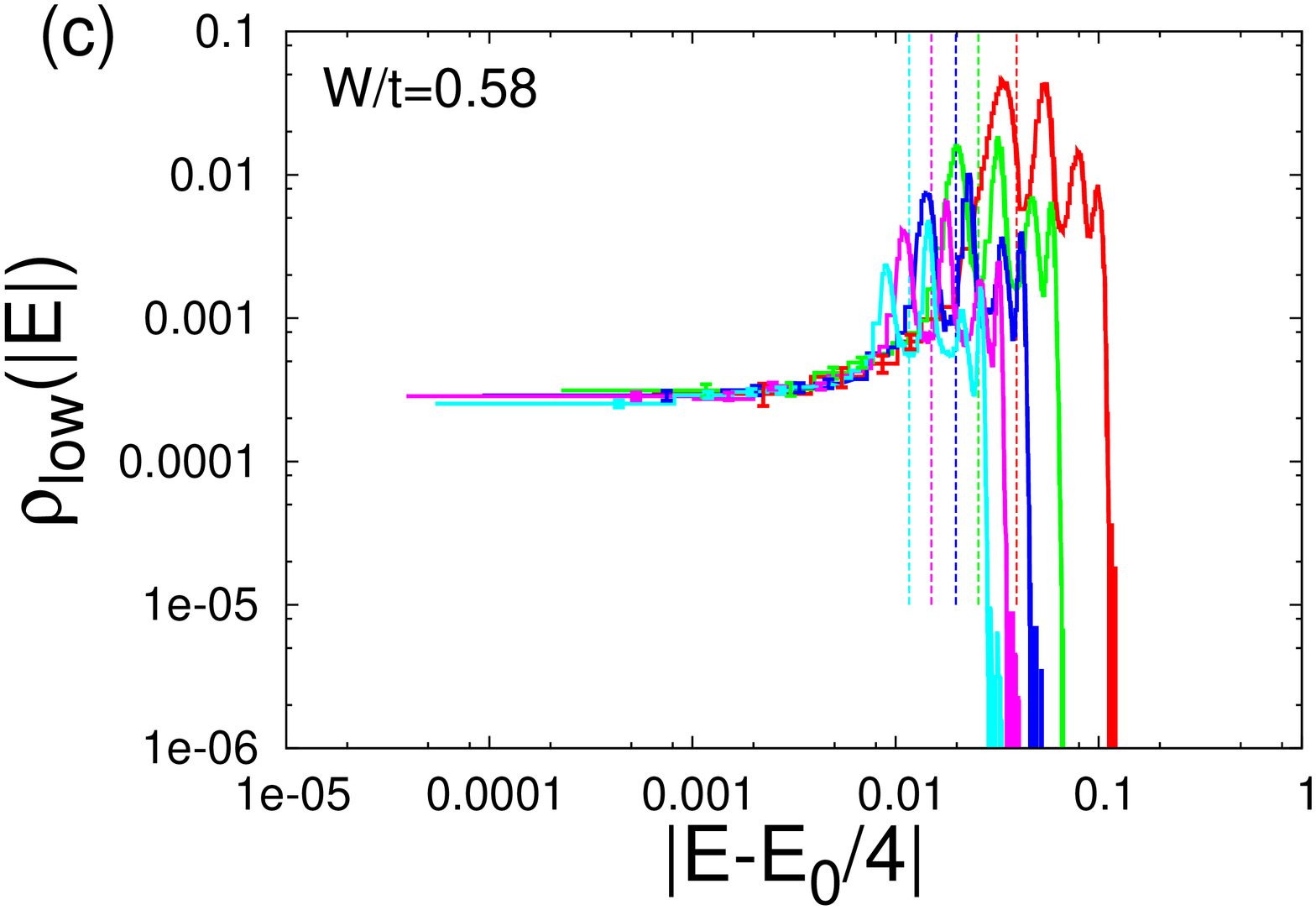}
\end{minipage}%
\caption{(color online) Low energy DOS computed using Lanczos on $(H-E_0(W,L)/4)^2$ with four states kept, and a twist $\theta_x=\pi/3$ for $W=0.50t$ (a),
$W=0.54 t$ (b), and $W=0.58t$ (c).  The vertical dashed lines mark $E_{N_u}^*$ where for larger energies the DOS may be underestimated
due to this calculation not getting the fifth and higher states.
The label for each $L$ is the same for each plot.  We find a clear $L$-independent low energy tail, which we take as an estimate of $\rho(0)$.  We also find that the Dirac peaks remain and continue to sharpen up for increasing $L$ in this range of $L$ and $W$, as detailed in Figs. ~\ref{fig:E_vs_W} and ~\ref{fig:E_vs_L}.}
\label{fig:dos_lan}
\end{figure*}

We now turn to the system size dependence of the $|E|$-peaks.
For the linearly dispersing Dirac excitations we find that each peak's energy follows the leading $1/L$ dependence inherited from its clean limit behavior as shown in Fig.~\ref{fig:E_vs_L}(a), with an exponent $x$ from $E \sim 1/L^x$ that varies from $0.995$ to $1.010$ (see the inset of Fig.~\ref{fig:E_vs_L}(a)).  We find this behavior clearly up to a disorder strength of about $W\sim 0.62t$ (not shown), beyond which cleanly identifying the energy of the peak above the (relatively) large background is no longer possible.  We have also checked that this holds for the second peak, as well, up to $W\sim 0.62t$ (not shown).
The FWHM $\Gamma$ of the first peak is shown in Fig.~\ref{fig:E_vs_L}(b), which is well described by the perturbative result $\Gamma \sim 1/L^{2}$ only for very weak disorder strengths $W\le0.1t$, which is consistent with where $\Gamma$ deviates from the perturbative expectation ($\sim W^2$) as in Fig.~\ref{fig:E_vs_W}(b).  For $W>0.1t$ we find a systematic decrease of the exponent $x$ in $\Gamma \sim 1/L^x$.

Thus we have shown that in the semimetallic regime, most properties of the eigenstates that make up the low-energy Dirac peaks in the DOS are well described by treating the Dirac eigenstates perturbatively in the disorder strength.
These Dirac peaks in the spectrum survive up to a disorder strength $W\approx 0.6~t$.  The one property that is not well captured perturbatively is the dependence of the width of the Dirac peaks on $W$, which grows faster than $\sim W^2$ in most of the SM regime.

For larger disorder strengths the model enters the avoided quantum critical regime,
with a substantially nonzero $\rho(0)$, and there is no longer a clear distinction between the peak and the background contributions to the DOS.  In this regime the rare eigenstates (that make up the background) are no longer rare at all and become \emph{typical} eigenstates, as the background fills in between the peaks.  It is both compelling and consistent that in our microscopic study of these Dirac peaks we find that they are no longer clearly part of the excitation spectrum for disorder strengths $W> 0.62t$, where the earlier KPM study using periodic boundary conditions on this model found the diffusive metal regime. Thus, the SM-DM crossover behavior survives the existence of the rare regions although the SM-DM QCP itself is destroyed by the rare regions!

\begin{figure*}[t!]
\centering
\begin{minipage}{.33\textwidth}
  \centering
  \includegraphics[width=0.7\linewidth,angle=-90]{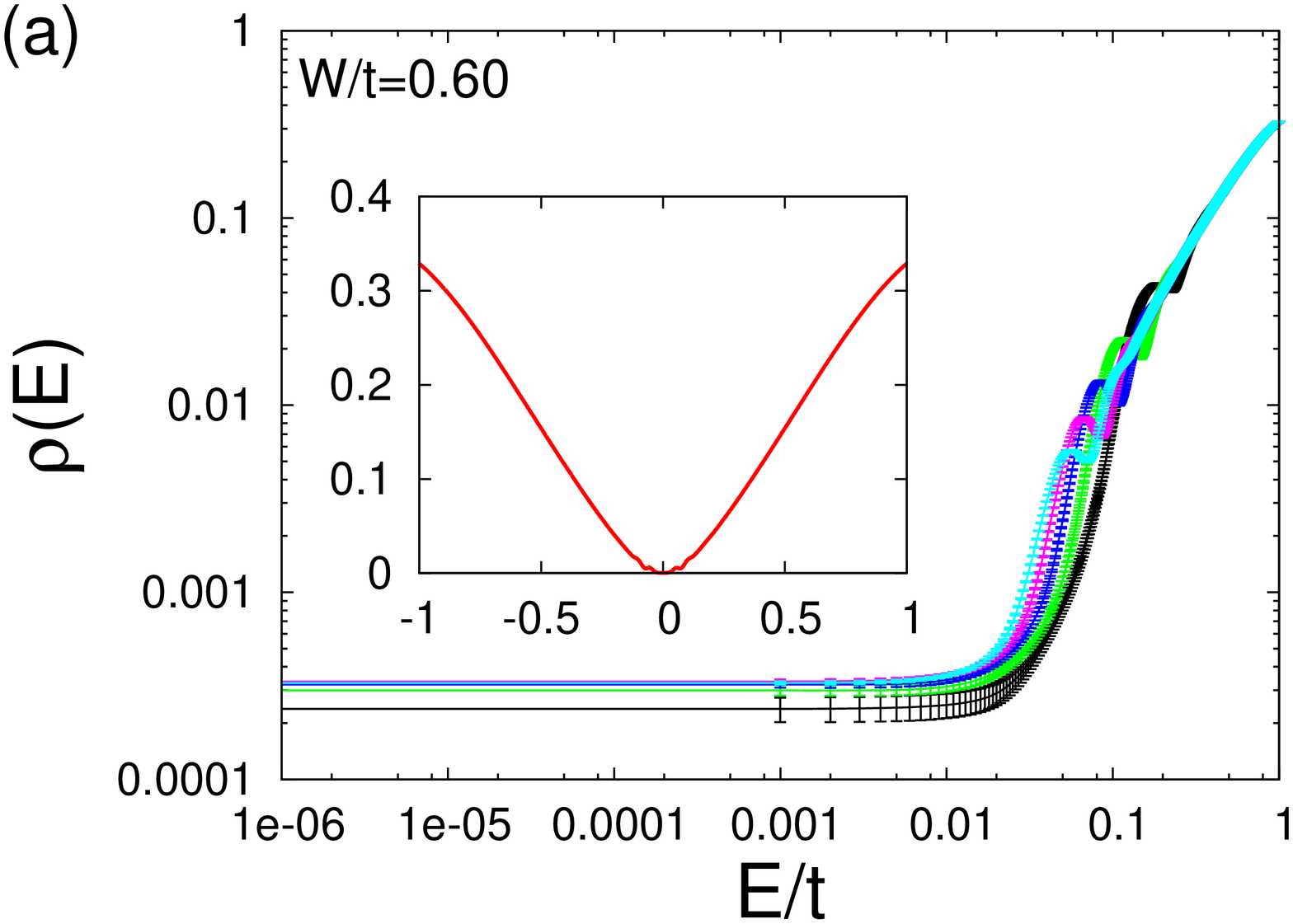}
\end{minipage}%
\begin{minipage}{.33\textwidth}
  \centering
  \includegraphics[width=0.7\linewidth,angle=-90]{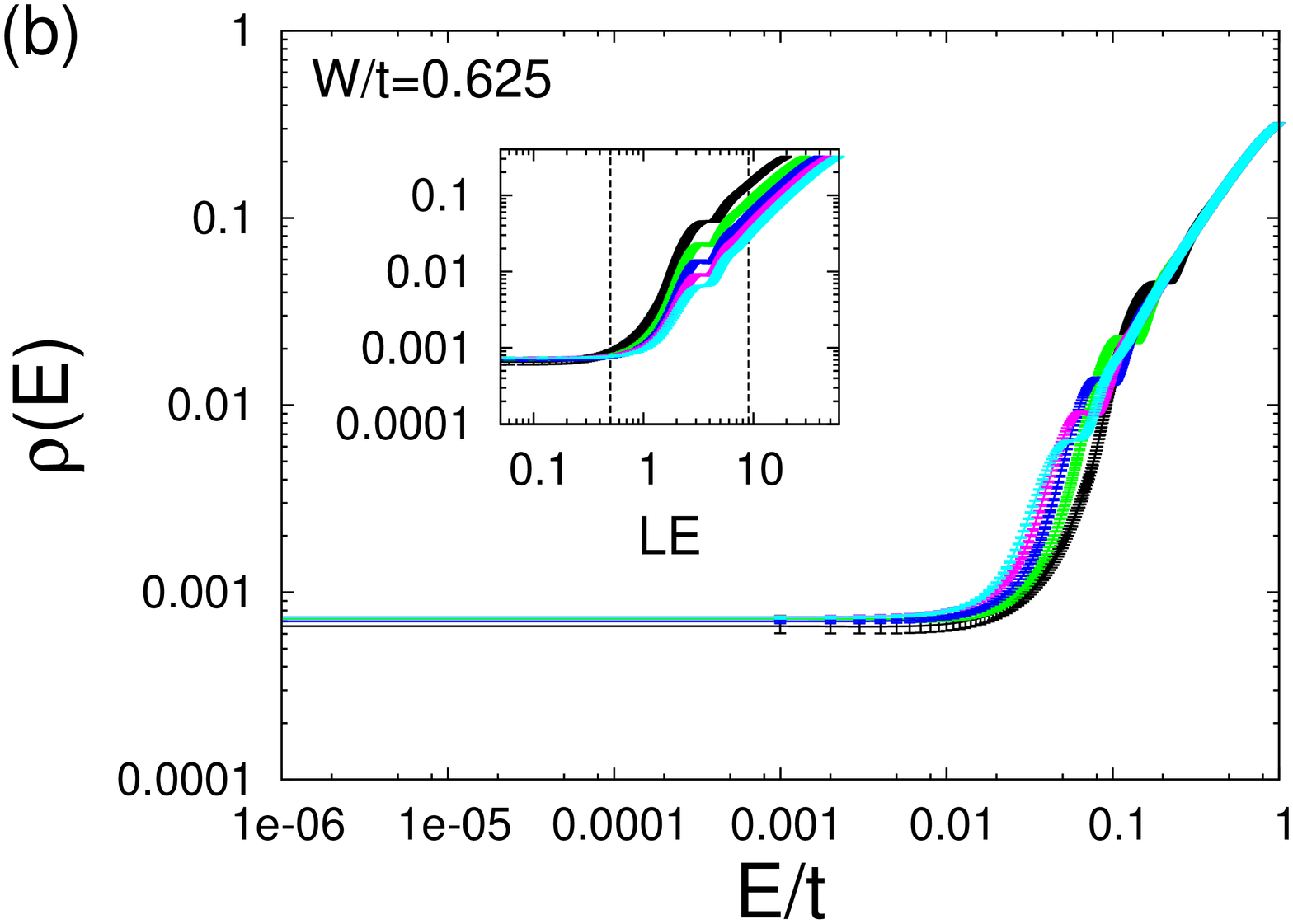}
\end{minipage}
\begin{minipage}{.33\textwidth}
  \centering
  \includegraphics[width=0.7\linewidth,angle=-90]{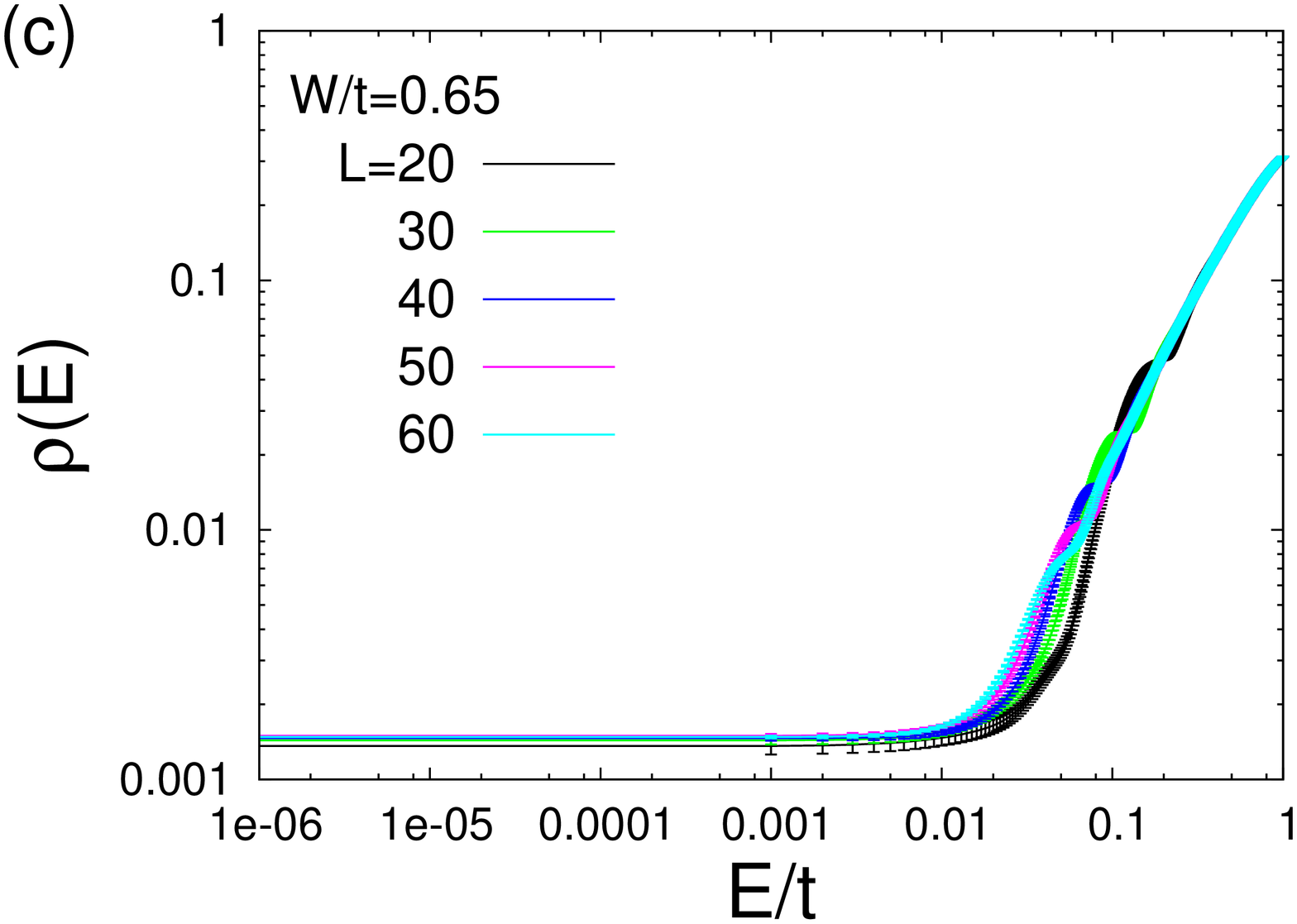}
\end{minipage}%
\caption{(color online) The full DOS $\rho(E)$ versus $E$ computed using the KPM on $H$ with a twist $\bm{\theta}=(\pi,\pi,\pi)$ for disorder
strengths $W/t=0.60$ (a), $W/t=0.625$ (b), and $W/t=0.65$ (c), with labels in (c) for each $L$ shared across all plots.  For each of these disorder
strengths we find a flat $L$-independent DOS for low energy, then the appearance of Dirac peaks or ``shoulders'' for intermediate energies,
and then an approximately power law DOS for larger energies.
Inset of (a): $\rho(E)$ versus $E$ for $W/t=0.60$ and $L=60$ showing the energy regime where approximately $\rho(E)\sim |E|$
that we suggest is the
quantum critical regime.  Inset of (b): $\rho(E)$ for $W/t=0.625$ versus $L E$ showing how the Dirac peaks line up as expected on this plot.}
\label{fig:dos_kpm}
\end{figure*}

\section{Density of States}
\label{sec:dos}

We are now in a position to estimate the rare eigenstate contribution to the zero-energy background DOS.  It is important to emphasize that
this is in contrast to the estimate of the Dirac peaks' contribution to $\rho(0)$ in Refs.~\cite{Pixley-2015,Pixley2015disorder}.
We first discuss our results for Lanczos on $(H-E_0(W,L)/4)^2$ with a twist $\theta_x = \pi/3$ and 10,000 disorder realizations for
each value of $W$ and $L$.
We then move onto the results using the KPM on $H$ with a twist of $\bm{\theta}=(\pi,\pi,\pi)$ and 1,000 disorder realizations.

The estimates of the low energy density of states from Lanczos on $(H-E_0/4)^2$ are shown in Fig.~\ref{fig:dos_lan} for disorder strengths $W/t=0.50, 0.54, 0.58$ which are all
in the semi-metal regime.
We find that the background contribution to the DOS develops an $L$-\emph{independent} low-energy tail, which is one of our main results.  Therefore this estimate of $\rho(0)$ is nonzero in the thermodynamic limit, albeit quite small for a weak disorder strength.  We also find that in this regime the Dirac peaks remain and continue to sharpen for increasing $L$; the energies and widths of them are shown in Figs.~\ref{fig:E_vs_W} and \ref{fig:E_vs_L} and discussed in section~\ref{sec:eigenstates}.  The orders of magnitude difference between the background and the Dirac-peak contributions to the DOS indicates the difficulty in trying to observe the rare region contribution to the DOS at zero energy without appropriately choosing the boundary conditions to explicitly separate these distinct contributions to the DOS.  We should note that these Lanczos estimates of the low energy DOS are actually of $\rho(E_0(L)/4)$, thus not at strictly zero energy.  But we find that in this very low energy range the DOS has only a weak energy dependence, so this is not a significant difference, especially compared to the roughly four order of magnitude range of the DOS as we vary $W$.

When estimating $\rho(0)$ using KPM, we would like to push the Dirac peaks as far away from zero energy as possible.  The Dirac peaks are broadened
both due to disorder and due to using finite $N_C$ in the KPM, and we want to minimize any contribution from this broadening
to the estimated $\rho(0)$.
This is achieved by using a twist of $\bm{\theta}=(\pi,\pi,\pi)$ in each direction and even $L$.  It is important to also remember that the KPM
introduces an artificial KPM background (as shown in Fig.~\ref{fig:dos_w=0.1}) which at small $W$ contaminates our estimate of the true background
DOS.  Therefore we cannot extend the KPM estimates of $\rho(0)$ to as small $W$ as we have for Lanczos.  As shown in Fig.~\ref{fig:dos_kpm}, using
the KPM for disorder strengths $W/t=0.600, 0.625, 0.650$, we find a flat low energy background contribution to the DOS that is
$L$-\emph{independent} and extends all the way to $E=0$.
Similar to the Lanczos data we still observe the Dirac peaks separating the smooth DOS at higher energy from the flat background, although for $W>0.6 t$ these peaks are being rounded out in to ``shoulders''.
For energies above these Dirac peak/shoulders we find that in this AQC regime
the DOS is close to the quantum critical (QC) form $\rho(E)\sim |E|$,
which is in agreement with the data in the absence of a twist~\cite{Pixley2015disorder}.  Thus, the crossover effects of the QCP survive, but the QCP itself does not.

\begin{figure*}[t!]
\centering
\begin{minipage}{.45\textwidth}
  \centering
  \includegraphics[width=0.75\linewidth,angle=-90]{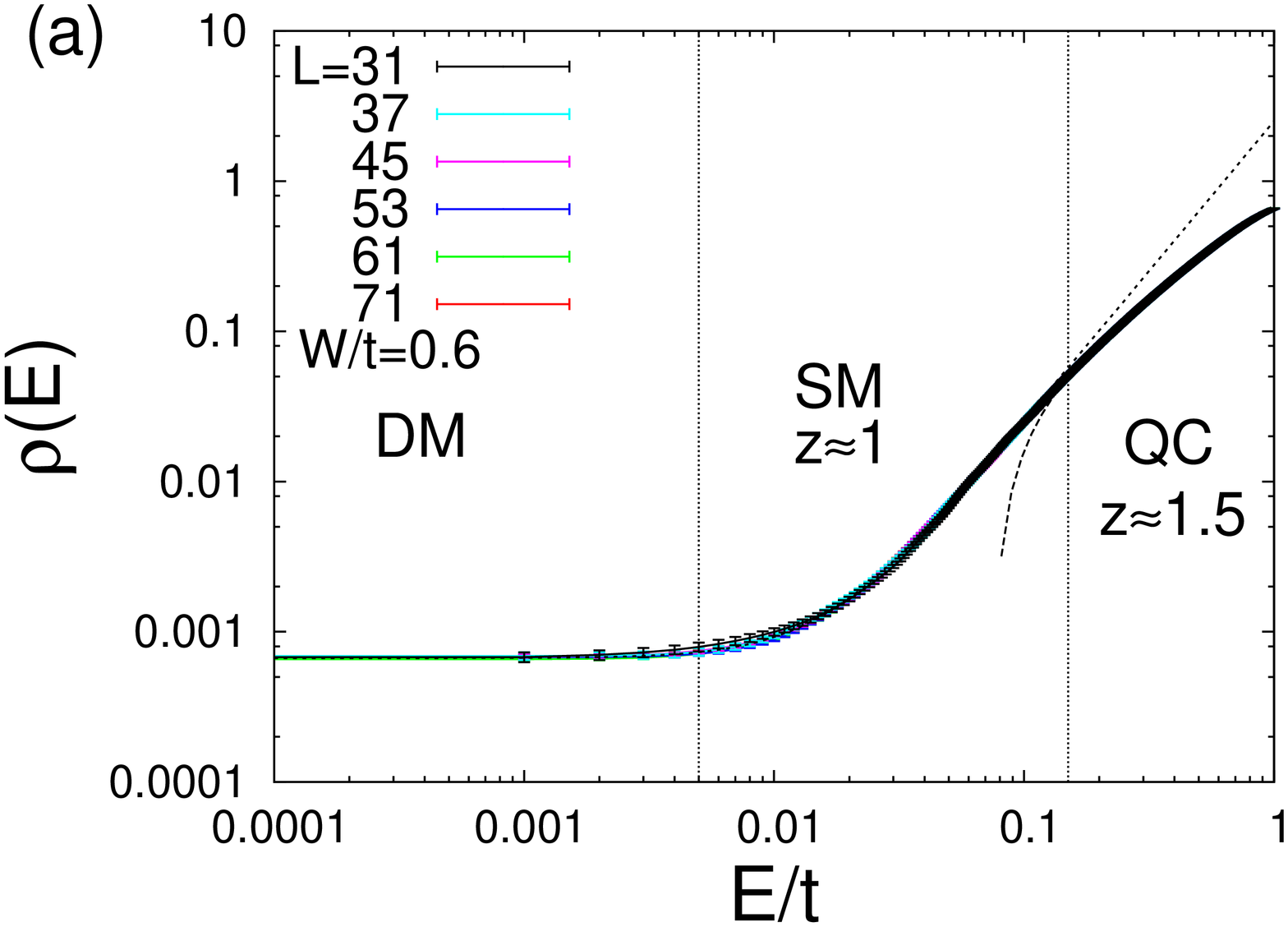}
\end{minipage}%
  \hspace{3mm}
\begin{minipage}{.45\textwidth}
  \centering
  \includegraphics[width=0.75\linewidth,angle=-90]{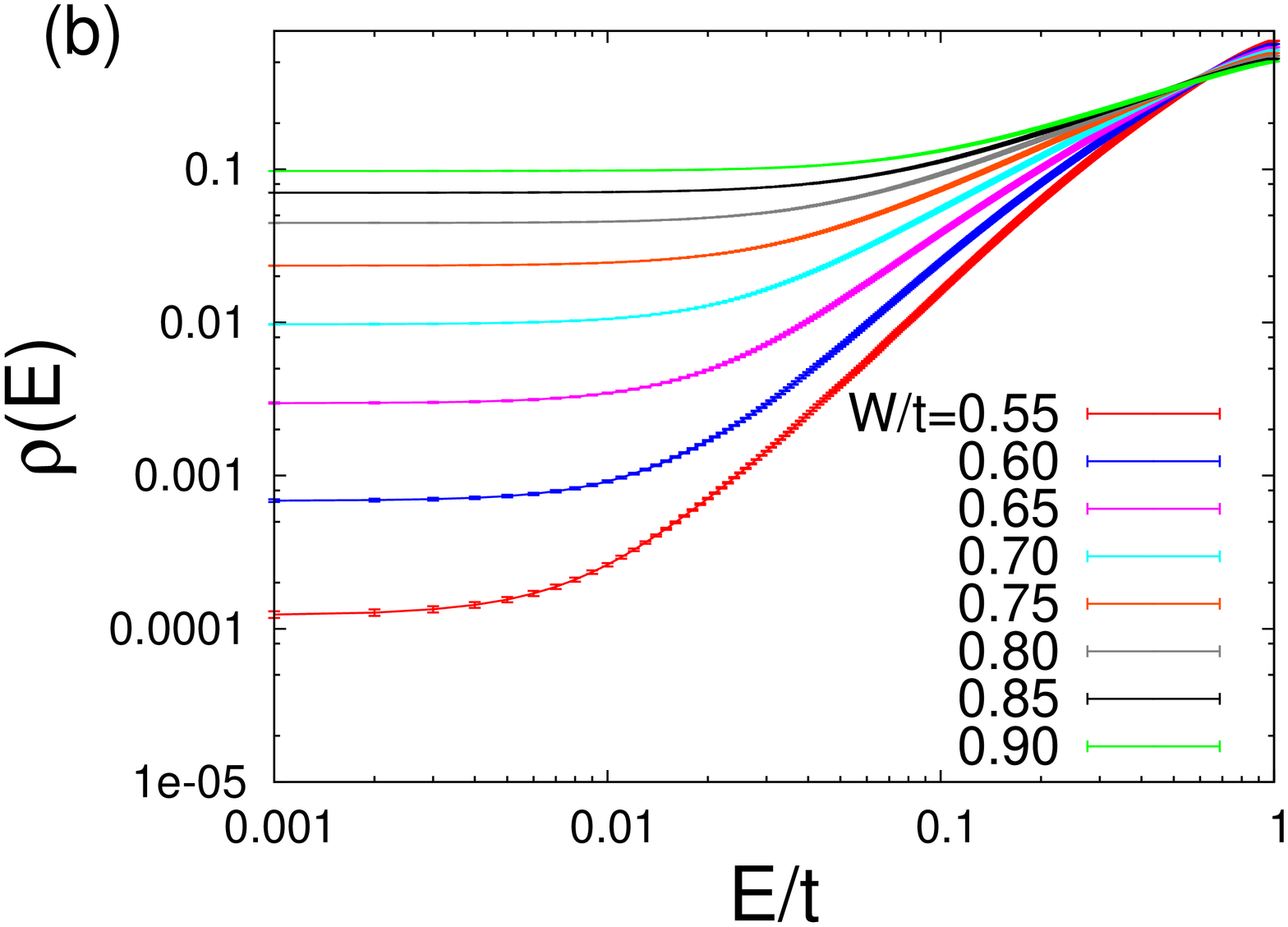}
\end{minipage}
%\newline
\centering
\begin{minipage}{.45\textwidth}
  \centering
  \includegraphics[width=0.75\linewidth,angle=-90]{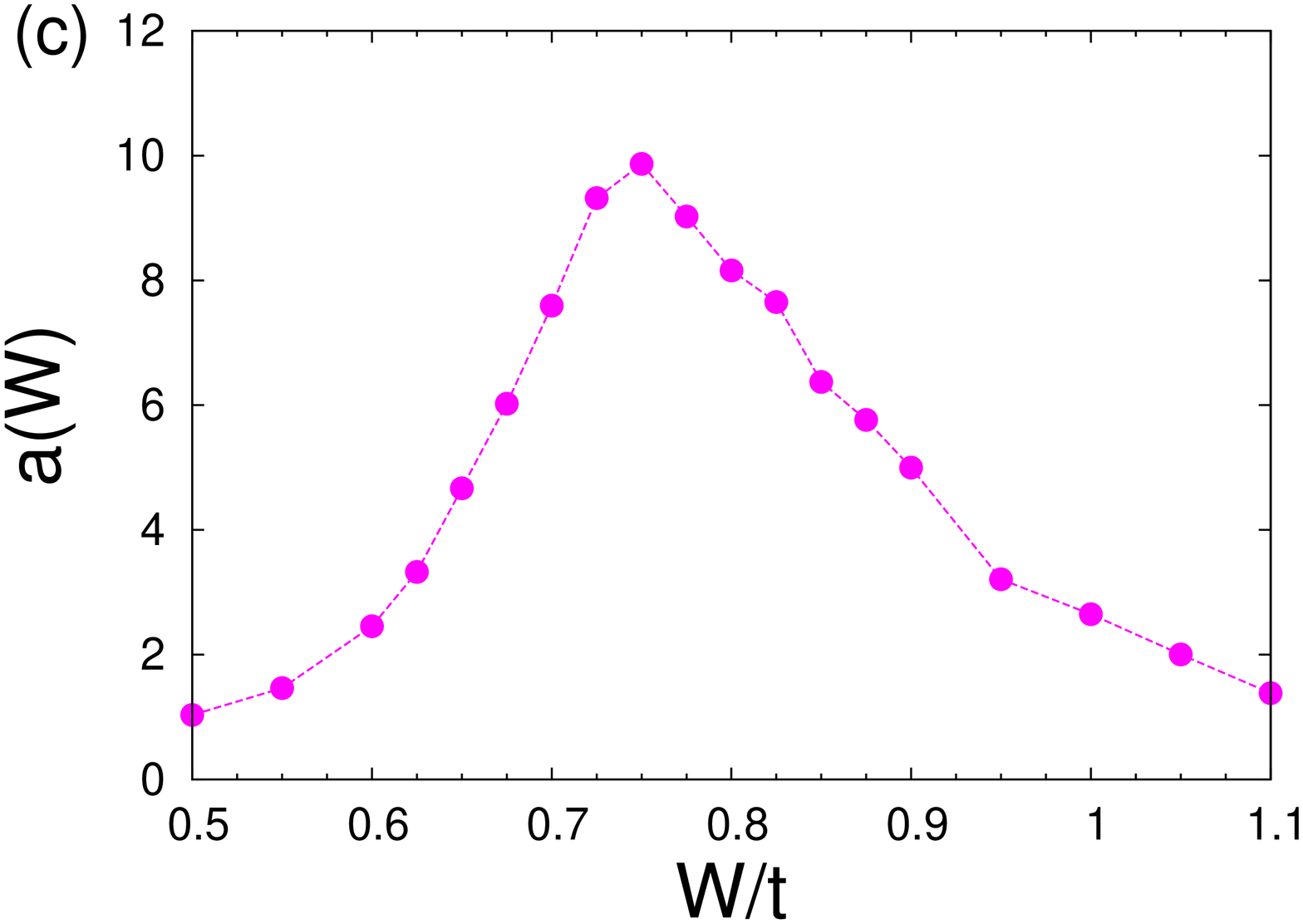}
\end{minipage}
  \hspace{3mm}
\begin{minipage}{.45\textwidth}
  \centering
 \includegraphics[width=1.1\linewidth]{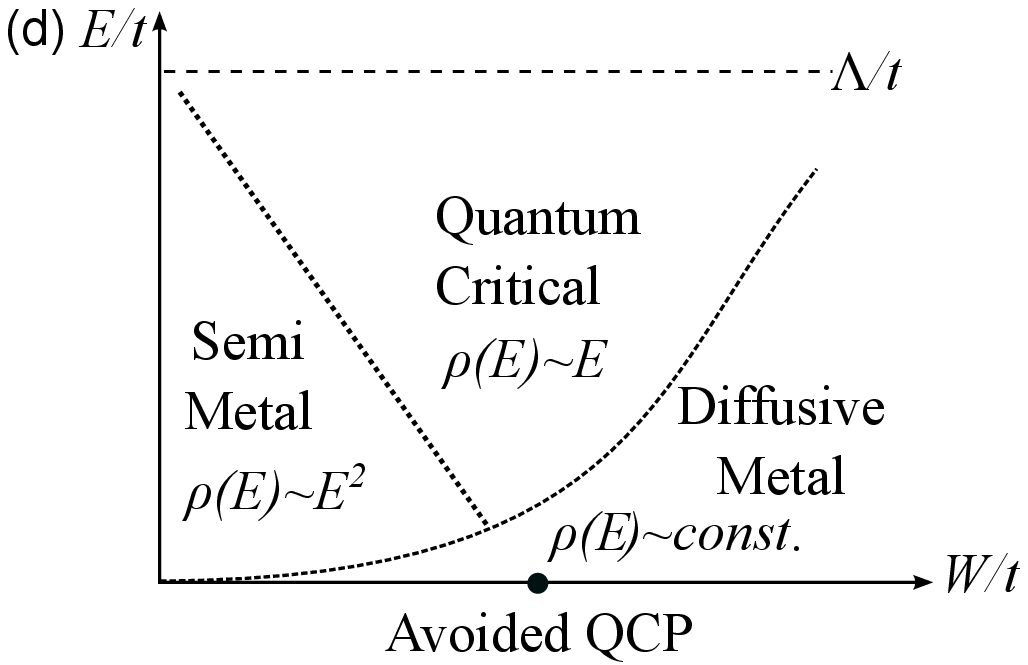}
\end{minipage}%
\caption{(color online) (a) DOS from the KPM with an expansion order $N_C=2048$ for $W/t=0.6$ averaged over the twist and disorder for $1,000$ realizations.  The data is well converged in system size for $L\ge 37$.  We find three distinct energy regimes separated by vertical (dotted) lines.  For sufficiently low energies the DOS is essentially flat with mainly diffusive excitations in the DM regime.  For intermediate energies (including the energy range where the Dirac peaks are in Fig.~\ref{fig:dos_kpm}) a fit of the DOS (thin dashed line) goes as $\rho(E) \approx \rho(0) + a|E|^x$ with $x=1.9\approx 2$ and $z=0.97\approx 1$, clearly identifying this as the SM regime.  For larger energies the DOS is fit to (thick dashed line) $\rho(E) = b+c|E|^y$, with $y=1.07$ and $z=1.45 \approx 1.5$, which is in good agreement with the expected scaling from the QCP.
(b) DOS from the KPM averaged over the twist and disorder for $1,000$ realizations with a linear system size $L=71$ for various values of $W$ displaying the cross over regimes in energy as a function of disorder. For increasing disorder the size (in $E/t$) of the DM regime increases while that of the SM regime decreases until for large enough disorder there is a direct cross over from the DM to the QC regime close to the avoided QCP at $W_c=0.75t$. Note that the value of $\rho(0)$ here is larger than for a twist of $\bm{\theta}=(\pi,\pi,\pi)$, as averaging over the twist does mix together the Dirac and rare states contributions to the zero energy DOS.
(c) Fit parameter $a(W)$ as a function of $W$ extracted from fitting $\rho(E) - \rho(0)$ to $a(W)|E|^2$ in the low energy limit from the KPM averaged over the twist and disorder for $1,000$ realizations with a linear system size $L=71$. 
We find $a(W)$ is a smooth function of $W$ and provides an accurate estimate of the avoided QCP $W_c = 0.75t$, thus we find the DOS is \emph{not} described by the combination of a smooth background and a critical part the critical point is sufficiently rounded out.
(d) Schematic crossover diagram as a function of energy and disorder strength.  Despite the existence of rare region effects we still find semimetal and quantum critical regimes exist, albeit at nonzero energies.  The quantum critical scaling regime is ``anchored'' by the avoided QCP and consistent with the perturbative RG analysis.}
\label{fig:dos_avg_twist}
\end{figure*}

The Dirac peaks in the SM regime are separated by $1/L$ (see inset of Fig.~\ref{fig:dos_kpm}(b)) and are a finite size effect, whereas the DOS is converged in $L$ both at high energy and near zero energy.  We can remove this finite size effect by averaging over the twist, which we do by generating a random twist vector $\bm{\theta}=(\theta_x,\theta_y,\theta_z)$ where each $\theta_i$ is a random twist for each disorder realization, uniformly distributed between $0$ and $\pi$; thus we average over the twist and disorder.  This is displayed in Fig.~\ref{fig:dos_avg_twist} (a) for $1,000$ disorder/twist realizations and a disorder strength $W/t = 0.6$.  For system sizes $L \ge 37$ we find the data is well converged in $L$ for \emph{all} energies, and as a result the data clearly displays three regimes in energy: the DM regime at the lowest energies, where the DOS does not depend on $E$; the SM regime at intermediate energies, where $\rho(E)\sim E^2$; and the QC regime at higher energy, where roughly $\rho(E)\sim |E|$. As shown in Fig.~\ref{fig:dos_avg_twist} (b): At smaller $W$, the QC regime disappears by moving up to near the cutoff.  At larger $W$, the size (in energy) of the DM regime grows and the SM regime disappears, leaving a direct DM to QC crossover, and at much higher $W$ the QC regime again disappears by moving up to near the cutoff.  It is important to emphasize that the DOS is always smooth through $E=0$ and $L$-independent for our largest values of $L$.
We can directly characterize this by expanding the DOS as $\rho(E,W) \approx \rho(0,W) +a(W)|E|^2+\dots$.
As shown in Fig.~\ref{fig:dos_avg_twist}(c), $a(W)$ rises smoothly developing a finite peak, with no divergence and no sign of any QC singularity at $E=0$.  The location of the peak value of $a(W)$ provides an estimate of the avoided QCP: we find $W_c \approx 0.75t$, which is consistent with the cross over regime in energy in Fig.~\ref{fig:dos_avg_twist} (b).  The implications of this are twofold: First, estimating the location of the QCP based on an apparent vanishing of $\rho(0)$ actually underestimates $W_c$ because (as we have shown) $\rho(0)$ is always non-zero.  Second, there is no indication at all of any singularity in the DOS in this system at any critical value of $W$ and the QCP is clearly rounded out (``avoided'').  This establishes that our numerical results are inconsistent with the DOS being expressed as a sum of two independent terms, a singular one arising from the quantum criticality and a smooth background contribution from rare regions.
This overall behavior is illustrated in the schematic disorder versus energy crossover diagram in Fig.~\ref{fig:dos_avg_twist}(d).

\begin{figure}[htb!]
  \centering
 \begin{minipage}{.5\textwidth}
  \centering
  \includegraphics[width=0.7\linewidth,angle=-90]{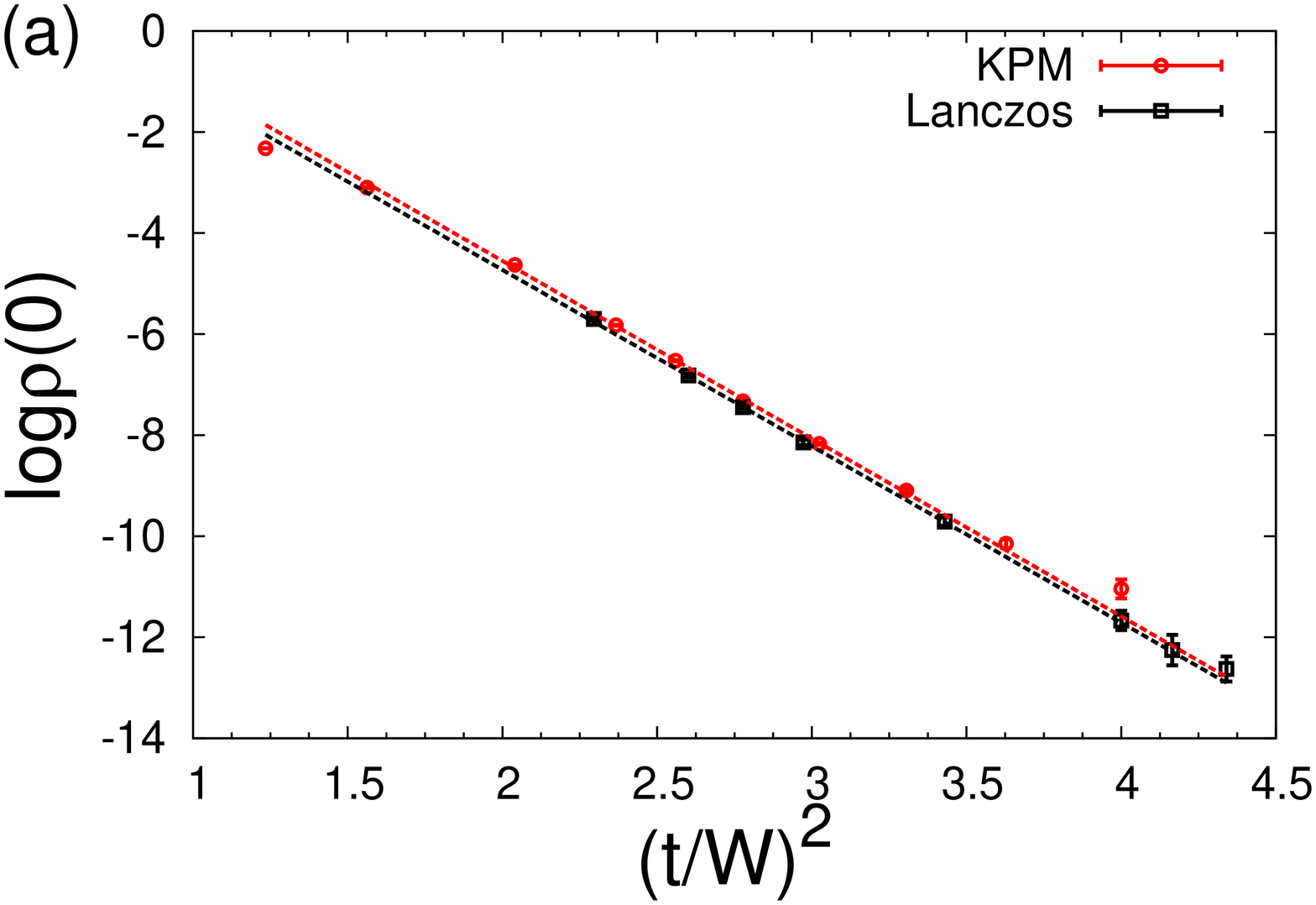}
  \end{minipage}
   \begin{minipage}{.5\textwidth}
  \centering
  \includegraphics[width=0.7\linewidth,angle=-90]{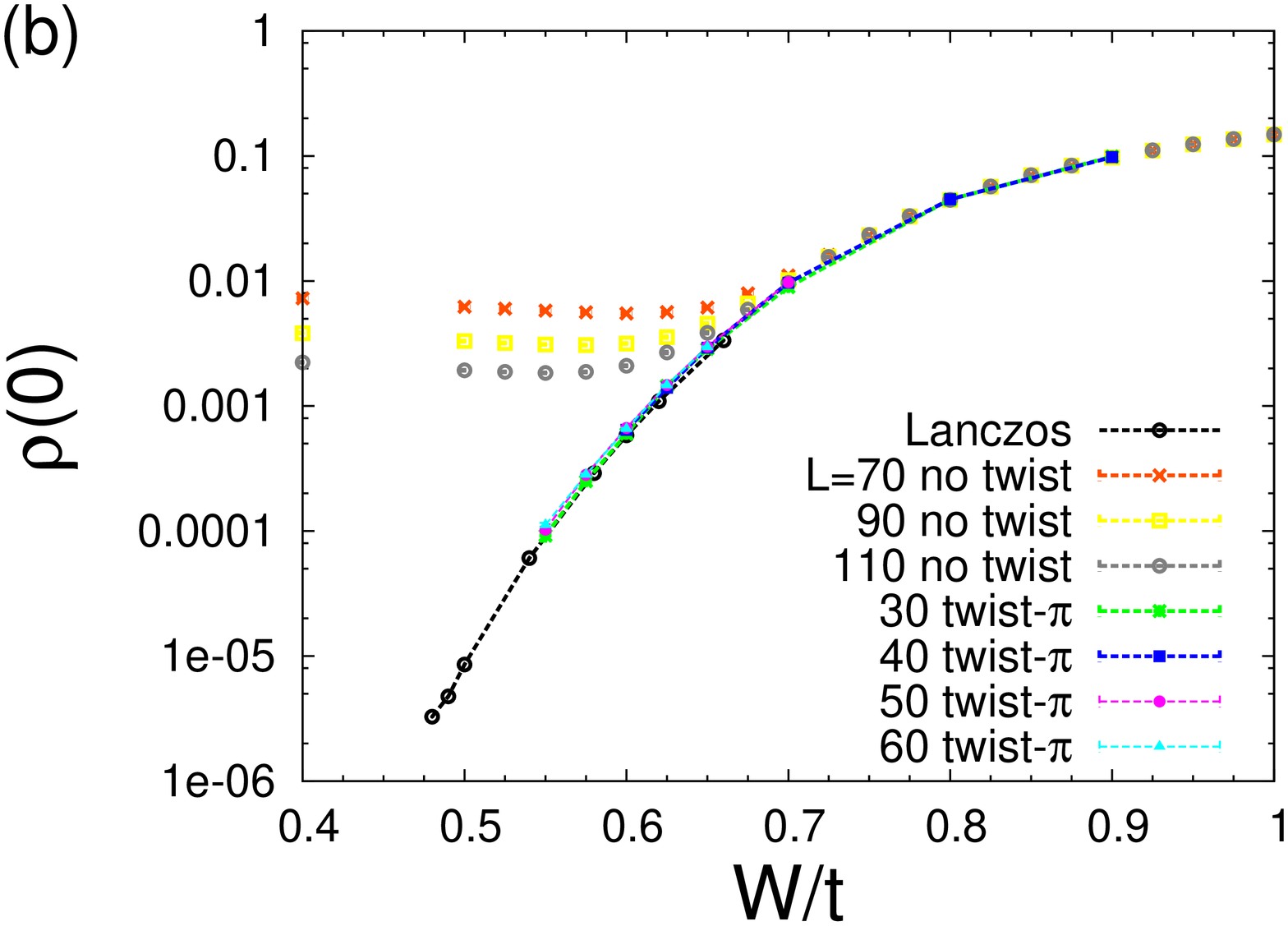}
  \end{minipage}
\caption{(color online) (a) Fit of the zero-energy background DOS to the rare-region form.  Log of $\rho(0)$ versus $(t/W)^2$ for the $L$-independent part of the background DOS computed from Lanczos and the KPM. (b) Various estimates of the zero-energy DOS from Lanczos and the KPM with and without a twist. The data for the KPM without a twist is reproduced from Ref.~\onlinecite{Pixley2015disorder}.}
\label{fig:dos_rr}
\end{figure}

Using the two estimates we now have from the $L$-independent part of the background DOS we fit our data to the rare-region form
\begin{equation}
\log \rho(0) \sim (t/W)^2,
\end{equation}
which fits remarkably well over four orders of magnitude of $\rho(0)$ going down to $W=0.48 t$ from Lanczos and up to $W=0.8 t$ for the KPM, with again no sign of any singular behavior at a critical value of $W$, as shown in Fig.~\ref{fig:dos_rr}(a).  We find good agreement between Lanczos and KPM estimates of $\rho(0)$ which implies that the background DOS is twist \emph{independent}. We find that the slopes of the fitted lines in Fig.~\ref{fig:dos_rr} (a) are in good agreement and match to within $\sim 1\%$ with a small offset between the two (about a $ 10 \%$ difference).  We attribute the systematic underestimate of $\rho(0)$ from Lanczos due to this method missing nearly degenerate eigenvalues that can arise from squaring the Hamiltonian. In addition, for small disorder strengths we see the KPM estimate starts to ``peel off'' systematically from the fitted line, we attribute this to the artificial KPM background in the DOS setting a lower bound to the value of $\rho(0)$ that we can accurately estimate with the KPM.

To conclude this section, we now discuss the various estimates of $\rho(0)$ using Lanczos, the KPM with no twist, and the KPM with a twist of
$\pi$ in all directions, as shown in Fig.~\ref{fig:dos_rr} (b).  Deep in the DM regime at high $W$, we expect that $\rho(0)$ should be completely
independent of the twist and $L$ (note this region cannot be reached with Lanczos as the spectrum is too dense), which is in good agreement with the numerics shown in Fig.~\ref{fig:dos_rr} (b).  For weaker disorder strengths the DOS from the KPM with a twist and the Lanczos estimates
exponentially decrease and remain $L$-independent, becoming orders of magnitude smaller than the $L$-dependent value of $\rho(0)$ without a twist.
As we illustrated in Fig.~\ref{fig:dos_w=0.1}(a), this is because a Dirac peak sits at $E=0$ if there is no twist, making this a very strong finite-size effect for
$\rho(0)$ in the semimetal regime.  The onset of this strong finite-size effect occurs near the avoided quantum critical point.  The avoided
quantum critical point also affects the DOS away from zero energy, giving it a quantum critical scaling of roughly $\rho(E)\sim |E|$ over a
range of energy. A crossover diagram summarizing this is
shown in Fig.~\ref{fig:dos_avg_twist} (d). We emphasize that although this crossover diagram is schematic, all aspects of it are obtained from our numerical results presented in this work.

\section{Discussion and Conclusion}
\label{sec:disc}

\begin{figure}[htb!]
  \centering
  \includegraphics[width=0.7\linewidth,angle=-90]{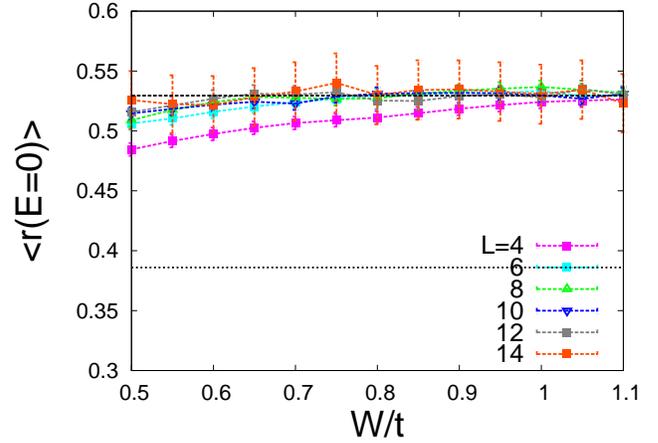}
\caption{(color online) Adjacent gap ratio focusing on the lowest four  $|E|$ eigenstates computed from exact diagonalization as a function of disorder for various system sizes. For system sizes $L=4,6,8,10$ we have used $10,000$ disorder realizations while $L=12,14$ we have considered $3,000$ and $1,000$ respectively. We find in the weak disorder regime the level statistics is GOE and therefore these low energy eigenstates \emph{are not} localized. The upper dashed line marks the expected GOE value $\langle r \rangle \approx 0.53$, while the lower dashed line marks the expected value of Poisson statistics $\langle r \rangle \approx 2\ln2-1$ (Ref.~\onlinecite{Oganesyan-2007}), for localized eigenvalues that have no level repulsion. Note that these results are consistent with the large disorder transition~\cite{Pixley-2015} (see Appendix) belonging to the orthogonal universality class.}
\label{fig:levelstats}
\end{figure}

Our results have demonstrated that in the semimetal regime the quasi-localized rare eigenstates live on a continuum of low energies that contribute a nonzero $L$ independent low energy DOS.  These rare eigenstates do not live in isolation and in principle there can actually be several per sample (here we have demonstrated a pair of these resulting in a bi-quasi-localized wavefunction as shown in Fig.~\ref{fig:wf_W=0.66}) with a non-zero tunneling matrix element that falls off with separation between the peaks ($r$) like $\sim1/r^2$.  All of these results are suggestive that these rare states are not fully localized.  Therefore we now turn to the low energy level statistics to directly address this question. We compute the adjacent gap ratio defined as
\begin{equation}
r_i = \frac{\mathrm{min}(\delta_i,\delta_{i+1})}{\mathrm{max}(\delta_i,\delta_{i+1})}
\end{equation}
where $\delta_i=E_{i+1}-E_i$ is the level spacing between neighboring energy eigenvalues. Here we focus on the center of the band and only compute $r_i$ for the lowest four $|E|$ eigenvalues through exact diagonalization on even $L$ samples with periodic boundary conditions, so that the level spacing will capture a mixture of both Dirac and rare eigenstates. Focusing on weak disorder where we can find the background DOS, as shown in Fig.~\ref{fig:levelstats} we find that the disordered average $\langle r \rangle \approx 0.53$ (to within numerical accuracy) for the biggest $L$ we have considered and therefore the level statistics satisfy the Gaussian orthogonal ensemble (GOE)~\cite{Oganesyan-2007}. We also find that $\langle r \rangle$ is unaffected by crossing the avoided QCP at $W\approx0.75t$. This establishes that the low energy eigenstates have a non-zero level repulsion and thus are not localized eigenstates. Therefore we can safely conclude that the quasi-localized eigenstates that fall off in a power law fashion are not localized, in strong contrast to the exponentially localized Lifshitz states (that live in a band gap or band edge).

In this work we have considered the effects of potential disorder on a three-dimensional
model that possesses an axial symmetry, which is pertinent to describing the bulk physics in various Dirac semimetals as well as
time-reversal-invariant Weyl systems.  We expect our results to be broadly applicable to models with the same symmetries
as $H_D$ or $H_W$. In this regard, Ref.~\onlinecite{Pixley2015disorder} found that $\rho(0)$ was numerically independent of tuning either potential, axial, or mass disorder for three-dimensional Dirac fermions, however now our work establishes that this is true with regards to the Dirac eigenstates only (as this was what was being computed in Ref.~\onlinecite{Pixley2015disorder}). Nonetheless, since the model with potential or axial disorder has a continuous axial symmetry, they can both be written in the form of $H_W$ while for mass disorder they cannot~\cite{Pixley-2015,Pixley2015disorder} and thus our results also apply to axial disorder. It will be interesting to see if this observation remains true with regards to the background (rare eigenstate contribution) to the DOS.  It will be exciting to explore disordered models with other symmetries, such as, for example, cases with disorder
that preserves particle-hole symmetry.  Perhaps there are other models where the semimetal is stable to disorder and a true phase transition
out of the semimetal occurs at some nonzero disorder strength.
This is, however, well beyond the scope of the current work where we consider the canonical (and by far the most studied) model of short-range potential disorder in the context of disorder driven SM-DM QCP, finding that the QCP does not exist (or becomes avoided) due to rare region effects, but the crossover effects of the QCP may exist in the quantum critical fan region.

Our results raise the question:  Is there a
field-theoretic description of these rare eigenstates and this avoided critical point?
It is now clear that the self-consistent Born approximation and perturbative RG are not capturing the crucial but nonperturbative effects of the rare disorder configurations.
It is important to mention that the analytic theory of these rare eigenstates has been obtained following a ``Lifshitz tail'' analysis and  Lifshitz tail eigenstates can also be taken into account within a field theory context and appear as an instanton configuration~\cite{Cardy-1978}.  It is therefore suggestive that the rare eigenstates that we have studied here will contribute some sort of non-pertubative instanton configuration that fundamentally changes the perturbative result (i.e. the self-consistent Born approximation) and the associated renormalization group analysis based on loop expansions.
Constructing this effective action incorporating the existing field together for both Dirac~\cite{Pixley2015disorder} and Weyl~\cite{Altland-2015,Altland2-2015} fermions should provide an effective theory for the avoided QCP and remains an important open question for the future.

The direct consequences of our findings on the non-trivial topological properties of clean Dirac and Weyl semi-metals such as the surface Fermi arc states~\cite{Wan-2011} and the axial anomaly~\cite{Adler-1969,Bell-1969,Nielsen-1983} is an interesting and open question. The absence of a bulk gap and the presence of these rare quasi-localized eigenstates may provide a scattering channel from surface arc states into the bulk endowing them with a non-zero quasiparticle lifetime and dissipative transport properties~\cite{Gorbar-2016} at sufficiently low energies. With regards to the axial anomaly that has been indirectly observed through a measurement of the longitudinal magneto resistance~\cite{Liang-2015,Huang2-2015,Zhang-2015}, we expect this does survive for reasonable transport time scales~\cite{Parameswaran-2014} as the the quasiparticle lifetime at low energies and weak disorder goes as $\tau\sim1/\rho(0)$ and will therefore be exponentially large in the strength of disorder. However, our results do establish that any perturbative treatment of the problem that is carried out at nonzero Fermi energy~\cite{Burkov-2015} cannot be extended down to $E=0$ (i.e. to the Dirac or Weyl cone) because of the existence of non-perturbative rare states. Lastly, the physics of the axial anomaly at larger fields in the quantum limit is unchanged by our findings because this is well described by quasi one dimensional dispersive states that host their own chiral anomaly in one dimension~\cite{Goswami-2015}. Both of these questions are sufficiently interesting and the effects of non-perturbative states upon them are sufficiently nuanced that they warrant their own separate study well beyond the scope of the present work.

It is interesting to compare our results for the avoided QCP with that of various strongly-correlated systems (such as heavy fermion metals~\cite{Hilbert-2007,Si-2014} or cuprate superconductors~\cite{Sachdev-2003}, where the evidence of a QCP is quite striking) where broken symmetry phases set in (such as superconductivity) and mask the zero temperature transition.  In these systems the quantum critical features are observed within the quantum critical fan and have a strong effect on finite temperature thermodynamic and transport properties.  With this in mind, and the schematic crossover diagram in Fig.~\ref{fig:dos_avg_twist} (b) that contains a quantum critical fan that is anchored by the avoided QCP, we still expect that if experiments on Dirac and Weyl semimetals can be tuned to the (zero energy) Dirac point thermodynamic signatures of the avoided QCP should show up in the crossover regime, e.g., a specific heat varying like $\sim T^2$. Upon lowering the temperature this power law will eventually cross over to $\sim T$ due to the rare regions masking the QCP.
Thus, in the current problem, the QCP is truly avoided (rather than `hidden') since there is no way, even as a matter of principle, to think of a situation to restore the QCP since disorder is the tuning parameter both for creating the QCP and for producing the rare regions destroying the QCP.  This is conceptually somewhat different from the situation with heavy fermions or cuprates where the origin of the superconductivity might be distinct from the origin of the QCP, at least as a matter of principle, so one can imagine suppressing the superconductivity (e.g. by applying a strong magnetic field) to restore the QCP.  In the Dirac-Weyl system, our current work definitively establishes that the disorder-driven SM-DM QCP does not exist as it appears to have been suppressed by the finite density of states contributed by the rare regions.  What does survive, however, is the crossover effect of the QCP which should produce effective scaling behavior provided one is at reasonably high energy (i.e. high temperature and/or high frequency). Such an apparent `effective scaling' behavior, numerically observed in many earlier theoretical studies, has led to the erroneous conclusion on the existence of a disorder-driven QCP in Dirac-Weyl systems, which our current work establishes as being nonexistent since it is avoided at the lowest energy (or the largest length) scale.

To conclude, we have studied the effects of rare eigenstates for Dirac and Weyl semimetals in the presence of potential disorder.  Using Lanczos (on $H^2$) and KPM (on $H$) with twisted boundary conditions we have established a systematic method to isolate and study the effects of rare regions in detail.  We have shown that for weak disorder the model under consideration possesses two classes of eigenstates.  Consistent with the perturbative irrelevance of the disorder, the first are Dirac eigenstates that are well described by perturbation theory in the random potential. These eigenstates disperse linearly with a wavefunction that is qualitatively consistent with a Dirac plane wave state weakly perturbed by the random potential.
The second class of eigenstates are the rare eigenstates that are very weakly dispersive and whose wave function is power-law quasi-localized near a site with strong disorder strength.
These eigenstates contribute a background DOS that extends all the way to zero energy and is exponentially small in the disorder strength $W$.  As a result of this non-zero DOS at zero energy, the expected semimetallic regime with DOS $\rho(E)\sim E^2$ only exists at energy scales above that set by the rare regions contribution to the DOS, and the apparent semimetal to diffusive metal QCP is avoided, pushing the quantum critical regime to nonzero energy.

\section{Acknowledgments}
We thank Pallab Goswami and Rahul Nandkishore for various discussions and collaborations on related work.  We also thank Victor Gurarie, Leo Radzihovsky, and Sergey Syzranov for discussions and for comments on a draft manuscript.  This work is partially supported by JQI-NSF-PFC, LPS-MPO-CMTC, and Microsoft Q (JHP and SDS). DAH is the Addie and Harold Broitman Member at I.A.S.
  JHP acknowledges the hospitality
of the Aspen Center for Physics (NSF Grant no. PHY-1066293) where some of this work was completed.
  We acknowledge the University of Maryland supercomputing resources (http://www.it.umd.edu/hpcc) made available in conducting the research reported in this paper. JHP would like to thank Zhentao Wang for help with the Lanczos package ARPACK.

\appendix
\section{Perturbation Theory}
\label{sec:appendix-A}
Here we determine the leading finite size and disorder effects from a perturbative analysis in the strength $W$ of the random potential.
For simplicity, we focus on the two-component model (3).  We use odd $L$ and include twisted boundary conditions such that the system has no
degeneracies at zero disorder, where the eigenenergies and eigenfunctions are
\begin{eqnarray}
E_{\bk,\pm}^{(0)} &=& \pm t\sqrt{\sin^2{k_x}+\sin^2{k_y}+\sin^2{k_z}}
\\
\psi_{\bk,\pm}^{(0)}(\br) &=& \frac{1}{\sqrt{L^3}}e^{i \bk \cdot \br}\phi_{\bk}^{\pm}
\\
\bk&=&\bk_0+2\pi(l,m,n)/L
\end{eqnarray}
where the $\phi_{\bk}^{\pm}$ are normalized two-component spinors,
$\bk_0$ is the wavenumber allowed by the twisted boundary conditions that is closest to zero, and $l$, $m$, $n$ are integers.

\subsection{First order correction to the eigenfunctions}
We have chosen the random potential $\tilde V(\br)$ to always sum to zero over all sites, so since there are no degeneracies at zero disorder the first order corrections to the eigenenergies all vanish.
The first order contribution to the eigenfunction $\psi_{\bk,\pm} (\br) =   \psi^{(0)}_{\bk,\pm}(\br) + \psi^{(1)}_{\bk,\pm}(\br)+...$ is given by
\begin{equation}
\psi^{(1)}_{\bk,\pm}(\br) = \sum_{\bq \neq \bk,s}\psi^{(0)}_{\bq,s}(\br) \frac{\sum_{\bR} \tilde V(\bR) (\psi^{(0)}_{\bq,s}(\bR))^{\dagger}\psi_{\bk,\pm}^{(0)}(\bR) }{E_{\bk,\pm}^{(0)}-E_{\bq,s}^{(0)}} ~,
\end{equation}
where $s$ is summed over $+$ and $-$.  For nonzero $W$, this sum is infrared divergent in the limit of large $L$ at energies away from the Dirac point, due to the small energy denominators.  This reflects the fact that the mean free path is finite away from the Dirac energy, and those eigenfunctions are highly changed by the scattering when $L$ exceeds the mean free path.  However, if we look at the eigenstates with energies closest to the Dirac point, as we do in this paper, then the vanishing of the DOS as $\sim E^2$ suppresses this infrared divergence and the sum is instead dominated by typical other states with energies far from $E_{\bk,\pm}^{(0)}$.  As a result, the first-order correction to these eigenfunctions is random with only short-range correlations and a relative magnitude that is $\sim W$ and independent of $L$ at large $L$.

\subsection{Second order correction to the eigenenergies}
The second order contribution to the eigenenergy is given by
\begin{eqnarray}
E^{(2)}_{\bk,\pm} &=& \sum_{\bq \neq \bk,s} \frac{|\sum_{\bR} \tilde V(\bR)(\psi^{(0)}_{\bq,s}(\bR))^{\dagger}\psi_{\bk,\pm}(\bR)|^2 }{E_{\bk,\pm}^{(0)}-E_{\bq,s}^{(0)}} ~.
\end{eqnarray}
We are interested in states near the Dirac point, so let's look in particular at $E^{(2)}_{\bk_0,+}$.  The contributions from typical momenta $\bq$ that are far from the Dirac points cancel to leading order:  Let's look at the contributions from the 4 other states at momenta $\bq=\bk_0\pm\bf{Q}$.  These two momenta are very close to equal and opposite, and as a consequence the corresponding eigenspinors $\phi^+_{\bk_0\pm\bf{Q}}$ are nearly identical to $\phi^-_{\bk_0\mp\bf{Q}}$, the corresponding energy denominators are of opposite sign and of nearly equal magnitude and magnitudes of the matrix elements are nearly identical.  Thus the level repulsion from the higher energy states almost exactly cancels that from the lower energy states.  What remains from these 4 other states is a random energy shift of order $W^2/L^4$ that on average lowers this positive eigenenergy $E_{\bk_0,+}$ by $\sim W^2/L^4$.  When these are summed over all $L^3$ other momenta, this gives an average decrease in this eigenenergy by $\sim W^2/L$, which gives a decrease of the Fermi velocity by $\sim W^2$.  The random contribution summed over all these typical other states to the eigenenergy is smaller by a factor of $L^{3/2}$, so is $\sim W^2/L^{5/2}$.

The contribution to $E^{(2)}_{\bk_0,\pm}$ from other states that are nearby in energy is random and $\sim W^2/L^2$, so does not contribute to the shift of the Fermi velocity in the limit of large $L$, but does dominate the random energy shift from typical other states.  Thus, when averaged over samples, this ``Dirac peak'' has linewidth $\sim W^2/L^2$ at second order in $W$.

\begin{figure}[h!]
  \centering
 \begin{minipage}{.5\textwidth}
  \centering
  \includegraphics[width=0.7\linewidth,angle=-90]{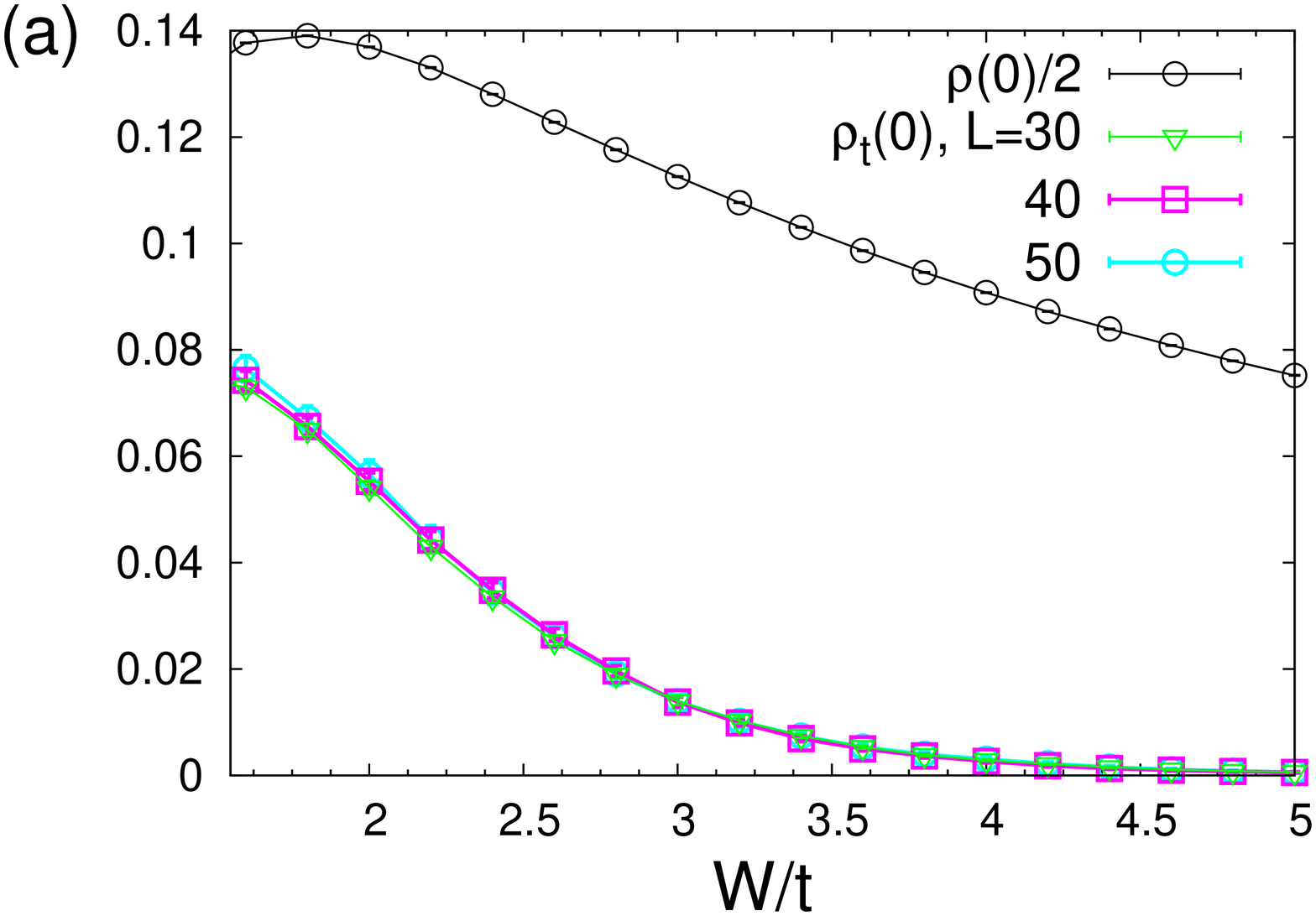}
  \end{minipage}
   \begin{minipage}{.5\textwidth}
  \centering
  \includegraphics[width=0.7\linewidth,angle=-90]{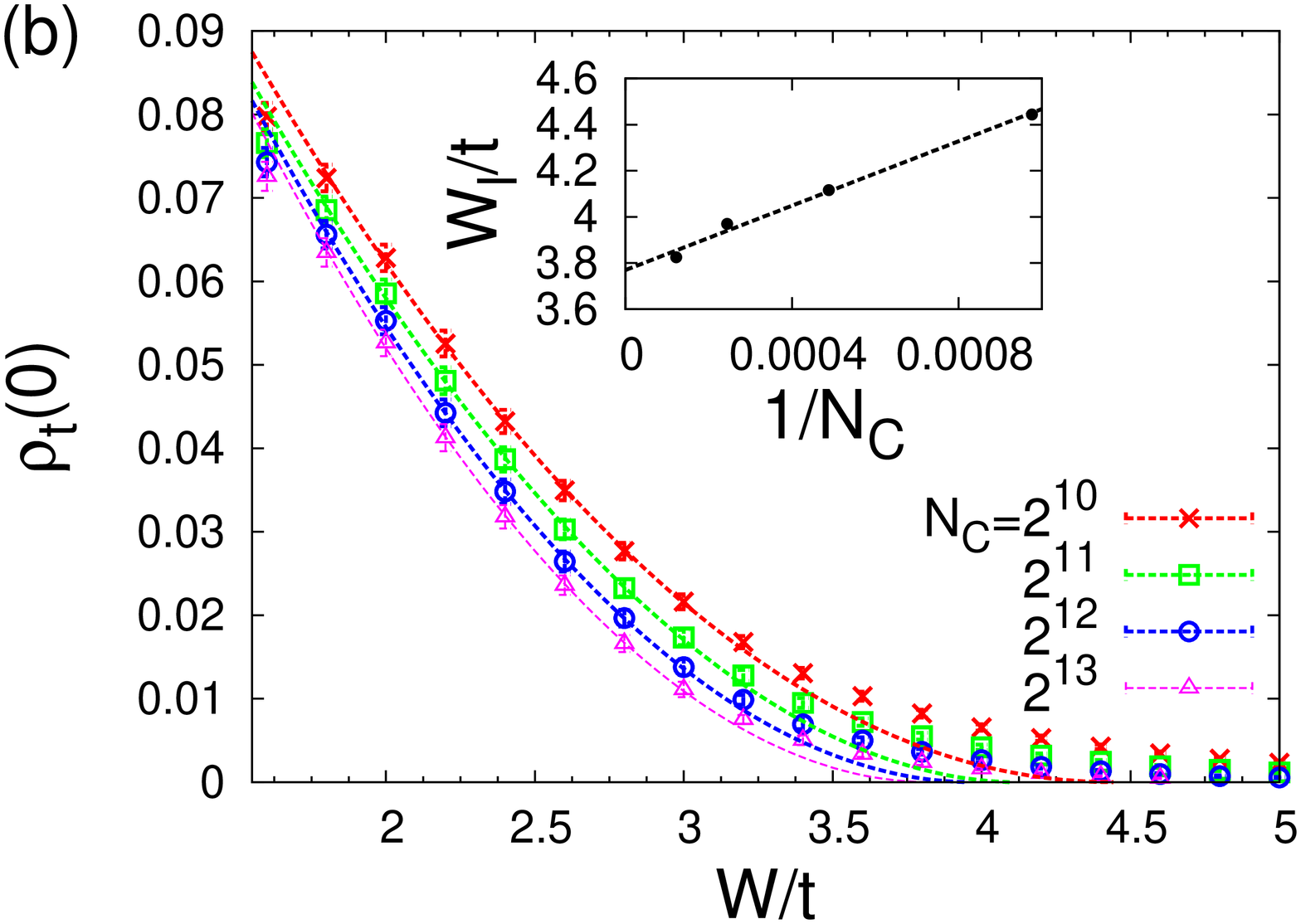}
  \end{minipage}
\caption{(color online) Localization transition at large disorder and $E=0$. (a) Averaged DOS averaged over the twist for $L=31$ with $1,000$ disorder realizations and the typical DOS with periodic boundary conditions for various systems sizes at a KPM expansion order $N_C=4096$ averaged over $4$ sites and $1,000$ disorder realizations, showing the data is well converged in system size for $L\ge 30$. (b) Determining the localization transition from where $\rho_t(0)$ extrapolates to zero in a power law fashion for $L=40$ as a function of $W$ for different KPM expansion orders. (Inset) Taking the limit $N_C\rightarrow \infty$ yields the localization transition at $W_l/t = 3.75 \pm 0.25$.} 
\label{fig:dos_t}
\end{figure}
\section{Localization transition at large disorder}
\label{sec:appendix-B}
In this section of the appendix we determine the location of the localization transition at large disorder, well away from the avoided QCP. For the model in Eq. (\ref{eqn:ham}) this transition has been studied in detail for the box distribution of disorder~\cite{Pixley-2015} but not for the gaussian distribution we have considered in the present manuscript. We determine the localization transition by first computing the local DOS $\rho_{i\alpha}(E)$ at a site $i$ and orbital $\alpha$ using the KPM~\cite{Weisse-2006} and then computing the typical DOS $\rho_t(E)$ from the geometric mean, these are defined as
\begin{eqnarray}
\rho_{i\alpha}(E) &=& \sum_{k,\beta}|\langle k,\beta |i, \alpha \rangle |^2\delta(E-E_{k\beta}),
\\
\rho_{t}(E) &=& \exp\left(\frac{1}{2N_s }\sum_{i=1}^{N_s}\sum_{\alpha=1}^2 \Big\langle \log \rho_{i\alpha}(E) \Big\rangle\right).
\end{eqnarray}
We only consider a few sites $N_s \ll V$ to improve the statistics and $\langle \dots \rangle$ denotes a disorder average. Due to the low DOS in the SM regime the typical DOS is not well suited to study the lack localization of the quasi localized rare states and therefore for this purpose we have used level statistics as shown in the main text in Fig.~\ref{fig:levelstats}. For large disorder the average DOS is sufficiently large and therefore the typical DOS is well behaved (see Fig.~\ref{fig:dos_t} (a)). We use periodic boundary conditions and even $L$ as for these large disorder strengths twisted boundary conditions have a negligible effect. We focus on the localization transition at $E=0$, as the mobility edge has been shown to be relatively standard~\cite{Pixley-2015}, i.e. it starts at the band edge and decreases in $|E|$ for increasing $W$. The results for large disorder and various systems sizes with a KPM expansion order of $N_C=4096$ are shown in Fig.~\ref{fig:dos_t}(a), we find $\rho_t(0)$ is well converged for $L\ge30$. To study the localization transition we fix the linear system size to $L=40$ and vary the KPM expansion order. By extrapolating $\rho_t(E=0)$ to zero in Fig.~\ref{fig:dos_t}(b) we find an estimate of the localization transition $W_l$ as a function of $N_C$, extrapolating to $N_C\rightarrow \infty$ yields an estimate of the localization transition at $W_l/t=3.75\pm0.25$. This places the standard Anderson localization transition for $E=0$ at a much larger disorder strength then the avoided QCP.

\bibliography{DSM_RR}

\end{document}